\documentclass[twocolumn]{aastex63}
\usepackage{amsmath}
\usepackage{txfonts}

\graphicspath{{./}{figure/}}

\begin{document}
\author{Kanon Nakazawa}
\affiliation{Department of Earth and Planetary Science, Tokyo Institute of Technology, Meguro, Tokyo 152-8551, Japan}

\author{Satoshi Okuzumi}
\affiliation{Department of Earth and Planetary Science, Tokyo Institute of Technology, Meguro, Tokyo 152-8551, Japan}

\author{Kosuke Kurosawa}
\affiliation{Planetary Exploration Research Center, Chiba Institute of Technology, Chiba 275-0016, Japan }

\author{Sunao Hasegawa}
\affiliation{Institute of Space and Astronautical Science, Japan Aerospace Exploration Agency, Kanagawa 252-5210, Japan}

\title{Modeling Early Clustering of Impact-induced Ejecta Particles Based on Laboratory and Numerical Experiments}

\shortauthors{Nakazawa et al.}

\correspondingauthor{Kanon Nakazawa}
\email{nakazawa.k.ai@m.titech.ac.jp}

\begin{abstract}
A projectile impact onto a granular target produces an ejecta curtain with heterogeneous material distribution. 
Understanding how the heterogeneous pattern forms is potentially important for understanding how crater rays form.
Previous studies predicted that the pattern formation is induced by inelastic collisions of ejecta particles in early stages of crater formation and is terminated by the ejecta's expanding motion. 
In this study, we test this prediction based on a hyper-velocity impact experiment together with N-body simulations where the trajectories of inelastically colliding granular particles are calculated.
Our laboratory experiment suggests that pattern formation is already completed on a timescale comparable to the geometrical expansion of the ejecta curtain, which is $\sim 10~{\rm \mu s}$ in our experiment. 
Our simulations confirm the previous prediction that the heterogeneous pattern grows through initial inelastic collisions of particle clusters and subsequent geometric expansion with no further cluster collisions.
Furthermore, to better understand the two-stage evolution of the mesh pattern, we construct a simple analytical model that assumes perfect coalescence of particle clusters upon collision. The model shows that the pattern formation is completed on the timescale of the system's expansion independently of the initial conditions. The model also reproduces the final size of the clusters observed in our simulations as a function of the initial conditions.
It is known that particles in the target are ejected at lower speeds with increased distance to the impact point. The difference in the ejection speed of the particles may result in the evolution of the mesh pattern into rays.
\end{abstract}


\section{Introduction}
\label{sec:intro}

Mutual collisions between planets, satellites, and minor bodies frequently occur in the solar system.
A hypervelocity impact of a small body onto a larger body leads to the production of an excavation flow that eventually opens a crater.
The materials in the excavation flow launch from the outer edge of a growing crater and form an inverted cone-like structure called an ejecta curtain.

Impact craters are often accompanied by radial streaks of ejecta called crater rays (e.g.~\citealp{1971Moon....2..263O}, \citealp{1989icgp.book.....M}, \citealp{HAWKE20041}).
Crater rays cause degradation of pre-existing craters, which may control crater equilibrium found for small lunar craters \citep{2019Icar..326...63M}.
Understanding how the morphology of crater rays is related to impact conditions may help us better understand how the size frequency distribution of small craters on planetary and lunar surfaces are determined.

While there are several studies characterizing crater rays on a variety of planetary bodies (e.g.~\citealp{1975JGR....80.2461T}, \citealp{2005Icar..176..351M}), 
only a few studies have addressed the physical mechanisms responsible for crater ray formation (e.g. \citealp{2012M&PS...47..262S}, \citealp{KADONO2015215, KADONO2020113590}, \citealp{2018PhRvL.120z4501S}). 
\citet{2012M&PS...47..262S} propose that interaction of impact-induced shock waves with pre-existing craters can produce crater rays.

More recently, \citet{KADONO2015215,KADONO2020113590} propose an alternative idea, which we focus on in this study, that crater rays are formed by an enhanced contrast in the particle number density in an ejecta curtain.
They conducted impact experiments onto granular targets and found that the resulting ejecta curtains have mesh-like patterns.
The observed mesh pattern simply expands with the ejecta curtain and suggested that the pattern formation had already been completed in an early stage of the ejecta curtain formation. They also performed numerical calculations and suggested that the mesh patterns  likely result from inelastic collisions between granular particles.
However, the previous work did not address when exactly the pattern formation occurred in the experiments and, more fundamentally, how the final size of the particle clusters constituting the mesh pattern is related to the initial condition of the ejecta curtain formation.
A better constraint on the timing of pattern formation is necessary for further investigating the processes of the pattern formation through future experiments.

In this study, we focus on the very early phase of ejecta curtain formation and study in detail how particle clustering proceeds and how it is terminated in an expanding curtain.
We present the results of a hyper-velocity impact experiment where we observed the development of a mesh-like pattern in the ejecta curtain on the timescale of the ejecta's expansion (Section \ref{sec:lab}). 
We also numerically simulate pattern formation of inelastically colliding granular particles with expanding systematic velocities (Section \ref{sec:sim}). To interpret the pattern formation observed in the simulations, we construct a simple physically-based model for clustering in an expanding particle system (Section \ref{sec:modeling}). 
We discuss our model's validity and the evolution of the mesh pattern into rays in Section \ref{sec:discussion}. A conclusion is presented in section \ref{sec:conclusion}.

\section{Laboratory Experiment}  
\label{sec:lab}

\subsection{Method}
\label{subsec:lab_method}
The purpose of our impact experiment is to observe pattern formation in a granular ejecta curtain immediately after an impact.
Imaging the early phase of crater formation using a video camera is generally hindered by self-emission from high-temperature vapor plumes. To avoid this issue, we use a monochromatic light source and a band pass filter, which enable us to image an ejecta curtain as early as 10 $\rm \mu s$ after the hyper-velocity impact of a millimeter-sized projectile. 
For comparison,  \citet{KADONO2020113590} conducted similar experiments but were only able to image the ejecta curtains $\sim 10 ~\rm ms$ after the impacts.

\begin{figure}[bt]
\gridline{\fig{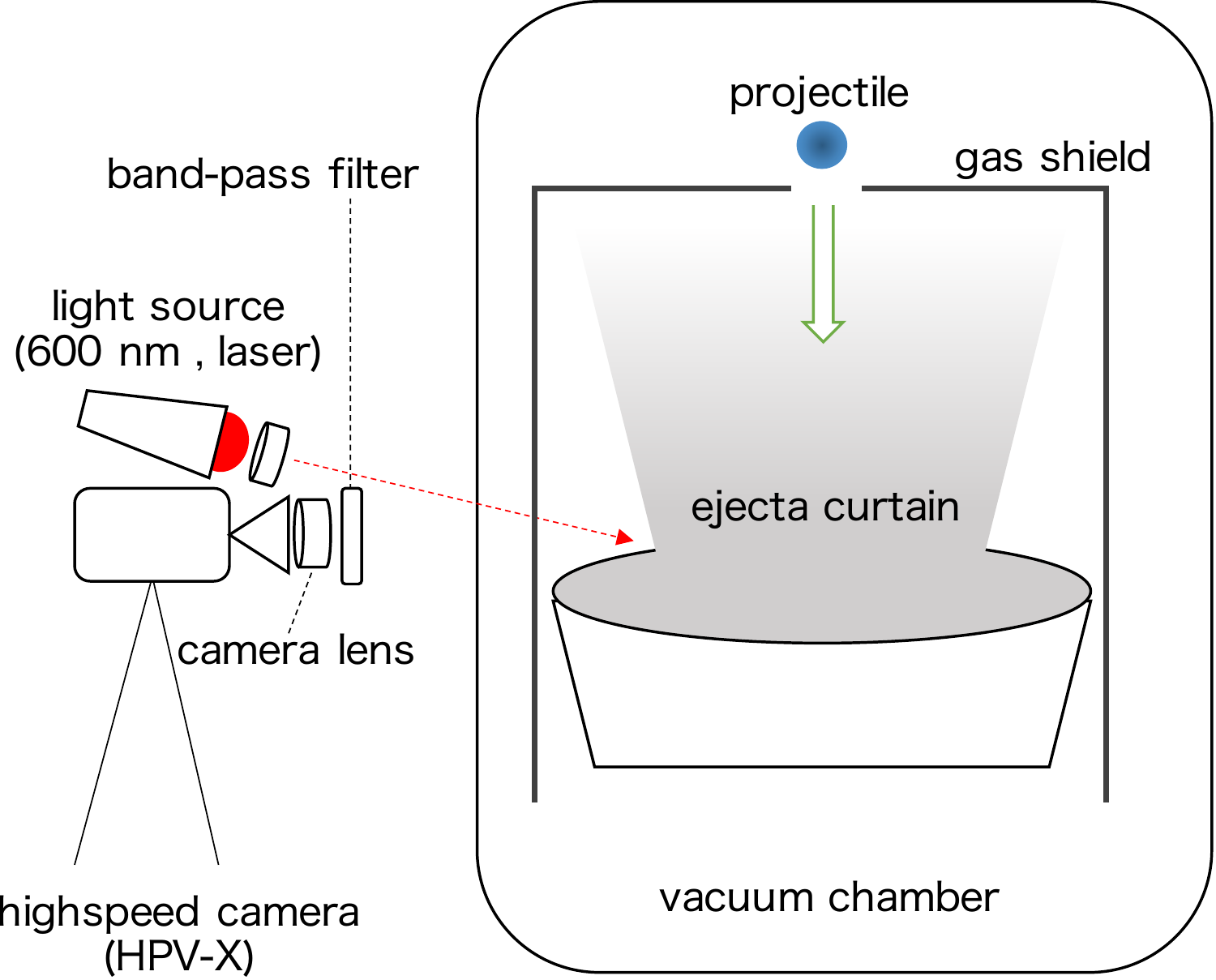}{0.45\textwidth}{}
          }
\caption{Illustration of the experimental setup.}
\label{fig:exp_concept}
\end{figure}

A schematic diagram of the experimental setup is shown in Figure~\ref{fig:exp_concept}. We used a vertical two-stage light gas gun at Institute of Space and Astronautical Science of Japan Aerospace Exploration Agency. A spherical polycarbonate projectile with a diameter of 4.8 mm was accelerated and impacted on the surface. We used zircon beads (FZB-400, the span of sizes is $30$ -- $63~\rm{\mu m}$) as granular targets. The median size of the beads was $50~\rm{\mu m}$. The impact velocity was set to  6.4~km~s$^{-1}$. The experimental chamber was evacuated down to 5 Pa prior to the shot. To cut off the propellant gas mixing associated with projectile acceleration, we surrounded the targets with a gas shield with only a narrow opening of several centimeters for the projectile. A high-speed video camera (Shimadzu, HPV-X) and a $640~\rm{nm}$ laser light source were used to observe the ejecta curtain.
The effects of self-luminous high-temperature plume on the curtain observation was minimized by mounting a band-pass filter corresponding to the laser wavelength in front of the camera lens. The spatial resolution was $110~\rm{\mu m}$ per pixel. We obtained a total of 128 frames with a frame rate of $1~\rm{\mu s}$. The frame rate is close to the characteristic time for the projectile penetration. Thus, we could observe the curtain growth as early as $\sim 10~{\rm \mu s}$ after the impact.

\subsection{Results}
\label{subsec:lab_result}
\begin{figure*}
    \gridline{\fig{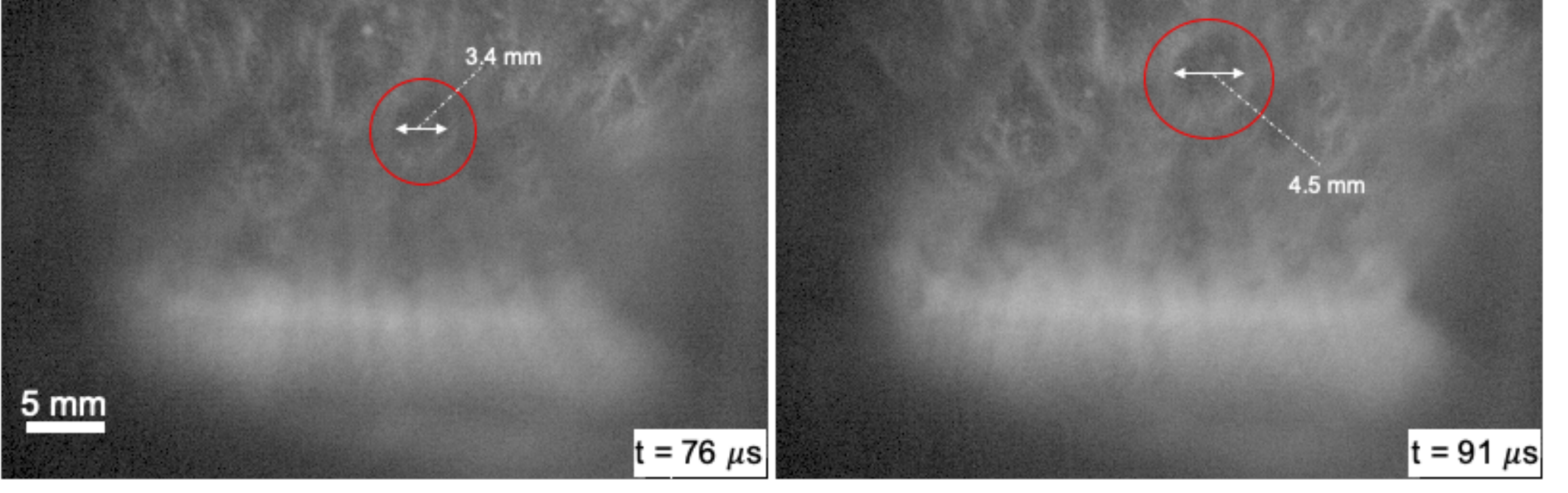}{0.9\textwidth}{}
          }
    \vspace{-1.0cm} \gridline{\fig{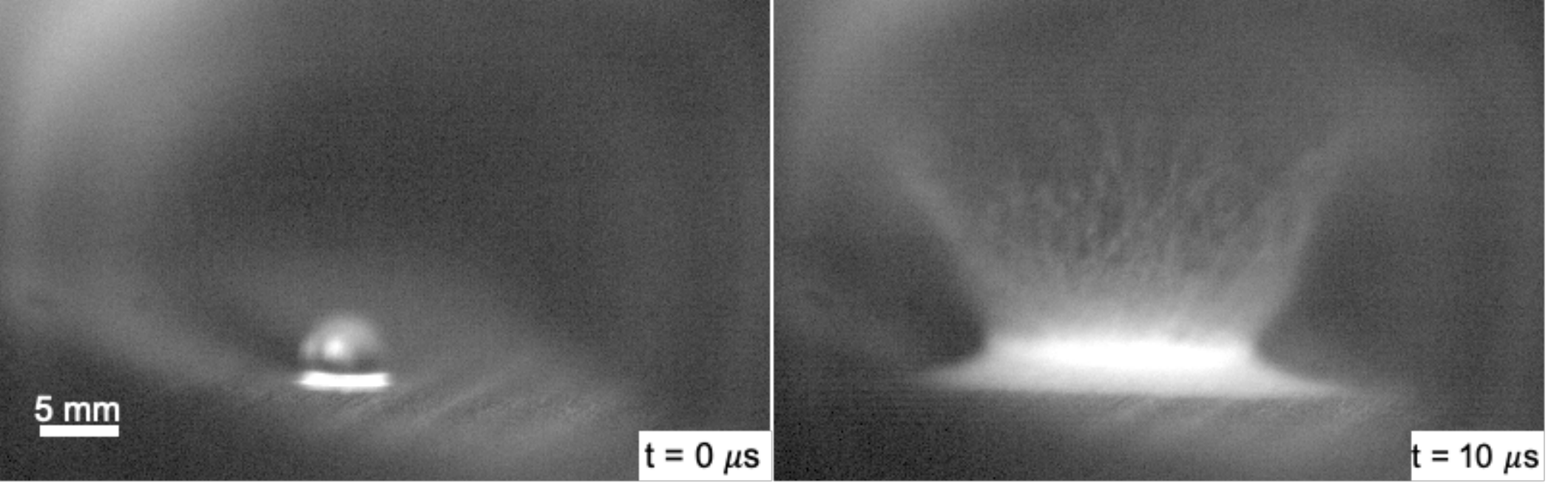}{0.9\textwidth}{}
          }
    \caption{Snapshots showing the earliest phase of  ejecta curtain formation from the laboratory impact experiment. The projectile impacts onto the granular target at $t = 0$ (upper left panel).  The red circles in the lower panels show how a pattern on the ejecta curtain expands geometrically with time.}
    \label{fig:experiments}
\end{figure*}

Figure~\ref{fig:experiments} shows the snapshots of the ejecta curtain within $100~\rm \mu s$ after the impact. The vertical and radial\footnote{We here adopt the cylindrical coordinate system centered on the impact point. The radial direction refers to the direction away from the vertical axis.} velocities of the ejecta in this early phase are estimated to be in the range of $\sim 100$--1000 $~\rm m~s^{-1}$. Because the ejecta immediately after the impact have a radial extent of $\sim 10~\rm mm$, they expand radially on a timescale of $\sim 10$--100 $\rm \mu s$.
Below, we call the timescale defined in this way the expansion timescale. 

We find that the ejecta curtain exhibits a mesh-like pattern already at $10~\rm{\mu s}$ after the impact (see the upper right panel of Figure~\ref{fig:experiments}). 
Moreover, the snapshots at 76 and 91 $\rm \mu s$ after the impact show that the pattern simply expands with the curtain (see the red circles in the lower panels of Figure~\ref{fig:experiments}).
\citet{KADONO2020113590} also observed the expansion of ejecta curtain patterns in their similar experiments, but at $\sim 10~\rm ms$ after impacts. 
Our experimental results presented here indicate that pattern formation occurs and is completed much earlier, on a timescale comparable to the expansion timescale of the ejecta curtain.

\section{Numerical Simulations}
\label{sec:sim}
In our laboratory experiment, we were unable to observe how the heterogeneous pattern developed during the very early phase of the ejecta curtain formation.
To better understand this process, we carry out N-body simulations where we calculate the motion and mutual inelastic collisions of granular particles in an expanding curtain. We particularly focus on the evolution of the particles' velocity dispersion and density pattern. To quantify the properties of the particle pattern, we regard the pattern as a collection of particle clusters and employ a cluster analysis.

\subsection{Method}
\label{subsec:sim_method}
Our simulations use the open-source N-body code  REBOUND~\citep{2012A&A...537A.128R}. We use IAS15~\citep{2015MNRAS.446.1424R} as an integrator. 
We approximate a portion of an ejecta curtain as a thin rectangular particle sheet and treat the motion of the ejecta particles in this portion in a two-dimensional Cartesian reference frame ($x,y$) comoving with the particles' center of mass, with the $x$ and $y$ directions corresponding to the directions parallel and perpendicular to the circumference of the cone-like curtain, respectively.
We neglect the gravity in this comoving frame assuming that the particles' center of mass follows a parabolic trajectory under the gravity. 
Because the particles in the laboratory frame move away from the impact point, the particle sheet expands in the $y$ direction. Including this expansion motion is essential here as we aim to see how the expansion of the curtain terminates the growth of particle clusters at late times. In our simulations, we mimic the expanding motion by adding a velocity proportional to $x$ to the $x$ component of the initial velocity of each particle (see below). In reality, the particle sheet would also expand in the direction perpendicular to the curtain's circumference if particles ejected earlier have higher ejection speeds (see Section~\ref{subsec:discussion_ray} for its potential implications for crater ray formation). For those cases, the $x$ axis in our simulations should be regarded as the axis along which the particle system expands faster.

The particles are initially placed in a square area of side length $L_0$. Before we set the initial velocities of the particles, we allow the particles to repel elastically in a periodic box to remove initial particle overlap.
Each particle (labeled by the subscript $j$) is given an initial velocity consisting of random and expansion components,
\begin{equation}
    (v_{jx},v_{jy})_{t = 0}=(v_{\mathrm{rand} , jx}+v_{\mathrm{exp},j},v_{\mathrm{rand} , jy}) ~,
\end{equation}
\begin{equation}
    v_{{\rm exp},j} = \Omega x_j~,
\end{equation}
where $t$ is time, $v_{{\rm rand},jx}$ and $v_{{\rm rand},jy}$ are random numbers, $v_{{\rm exp},j}$ is the initial expansion velocity, and $\Omega$ is the rate of expansion.
With this initial condition, the particles expand in the $x$ direction and also mutually collide nearly isotropically. 
The random velocity components are sampled from a uniform distribution ranging from $-v_{\rm rand,max}$ to $v_{\rm rand,max}$, where $v_{\rm rand,max}$ is a constant. 
Whenever two particles collide, we decompose their velocities into the center-of-mass and relative velocity components and change the latter as
\begin{equation}\label{eq:e_collision}
     v'_{\rm rel,n} = -ev_{\rm rel,n}, \qquad 
     v'_{\rm rel,t} = v_{\rm rel,t},
\end{equation}
where $v_{\rm rel,n}$ and $v_{\rm rel,t}$ are the normal and tangential  components of the relative velocity before the collision, respectively, $v'_{\rm rel,n}$ and $v'_{\rm rel,t}$ are those after the collision, and $e$ is the coefficient of restitution. We ignore the rotation of the particles by assuming that the particles exert no torque on each other upon collision. In our simulations, we set $e$ as a constant.
We carry out 10 independent simulation runs with 10 different sets of initial particle random velocities but with the same initial particle configuration.

\begin{deluxetable}{lc}
\label{tb:parameter}
\tablecaption{Parameters Adopted in the Simulations}
\tablecolumns{2}
\tablewidth{0pt}
\tablehead{
\colhead{Parameter} &
\colhead{Value}
}
\startdata
    Total number of particles $N_{\rm tot}$ & 30000  \\
    Particle radius $r$ & 50 $\rm{\mu m}$ \\
    Expansion rate $\Omega$ &  $0.02~\rm{\mu s^{-1}}$  \\
    Maximum random speed $v_{\rm rand,max}$ &  $40~\rm{m~s^{-1}}$ \\ 
    Restitution coefficient $e$ & 0.6 \\
    Initial box size $L_{0}$ & 20~mm  
\enddata
\end{deluxetable}

The values of the parameters adopted in the simulations  are listed in Table~\ref{tb:parameter}.
All particles in the simulations are assumed to be spherical and equal-sized, with radius $r$ being equal to those of the zircon beads used in the experiment presented in Section~\ref{sec:lab}.
Because we exert no external force on individual particles, our simulation results are independent of the particle mass. For this reason, we adopt the code units in which the particle mass is taken to be unity.
The expansion rate $\Omega$ is set to $0.02~\rm{\mu s}^{-1}$ so that the expansion velocity of the system, $\Omega L_0/2 = 200~\rm m~s^{-1}$, is comparable to the horizontal velocity of the ejecta curtain pattern observed in our laboratory experiment. The maximum random velocity is taken to be 20\% of the system expansion velocity.
The coefficient of restitution of our zircon particles are unknown, so we arbitrarily take the default value of $e$ to be 0.6 so that the pattern formed is visible enough to analyze. The validity of our choice of $e$ is discussed in Section \ref{sec:discussion}. 
\citet{KADONO2015215} show that particles of a lower restitution coefficient produce a clearer particle pattern through inelastic collisions. For completeness, we also present simulations for $e = 0.1$, 0.9, and 1.0.

The time step $\Delta t$ must be chosen to be smaller than the crossing times of the particles so that every particle collision can be detected. 
The particle crossing time is estimated as $2r/v_{\rm rel} \ga r/v_{\rm rand,max} \approx 1.3~\rm \mu s$, where $v_{\rm rel}$ is the relative velocity of the particles. We therefore set $\Delta t = 0.15~\rm{\mu s}$.

\subsection{Data Analysis}
\label{subsec:sim_analysis}
Particles lose their random kinetic energies through inelastic collisions. Since the particles also form clusters through the collisions, the random energy of the system and the size of the clusters should be correlated with each other. Below we describe how we quantify the two quantities.

As a diagnostic for the random motion of the particles, we define the root-mean-squared (RMS) particle velocity as 
\begin{equation}
\label{v_rms_sim}
v_{\rm{RMS}}=\langle(v_{jx} - v_{{\rm exp},j})^2 + v_{jy}^2\rangle^{1/2}~,
\end{equation}
where the brackets represent an average over all particles (again, each particle is labeled by $j$).
To reduce statistical errors, we also take the average of $v_{\rm RMS}$ over 10 runs. 
The masses of particle clusters are measured using a cluster analysis. In this analysis, all particles in a given snapshot of simulations are grouped into  clusters.
We adopt hierarchical clustering in which 
pairs of clusters (initially individual particles)  are combined successively.
Specifically, we use Ward's method~\citep{doi:10.1080/01621459.1963.10500845}, a commonly used hierarchical clustering algorithm that combines clusters so that the variance of the particle positions within the clusters
is minimized.
However, this algorithm alone does not tell us the {\it optimal} (i.e., most likely) number of clusters. To determine it, we introduce a quantity called the silhouette coefficient. For each particle $j$, its silhouette coefficient $s_j$ 
is defined as \citep{ROUSSEEUW198753}
\begin{equation}\label{eq:silhouette}
    s_j = \frac{b_j - a_j}{\mathrm{max} \{ a_j, b_j\}}~,
\end{equation}
where $a_j$ and $b_j$ are the average distances between particle $j$ and the other particles inside and outside the the cluster, respectively, to which it belongs. The higher the average silhouette coefficient $\langle s_j \rangle$ is, the better the clusters are separated.
For each snapshot of particle distribution, we pick up the set of clusters that give the highest $\langle s_j \rangle$, and calculate the mean and standard deviation of their masses. We also take averages of the means and standard deviations over 10 simulation runs.
The cluster analysis we employ is more naturally suited to our problem than Fourier analysis as used by \citet{KADONO2015215}, because our particle system has an open, expanding boundary and therefore the cluster distribution within it is nonuniform.

\subsection{Results}
\label{subsec:sim_result}
\begin{figure*}
\gridline{\fig{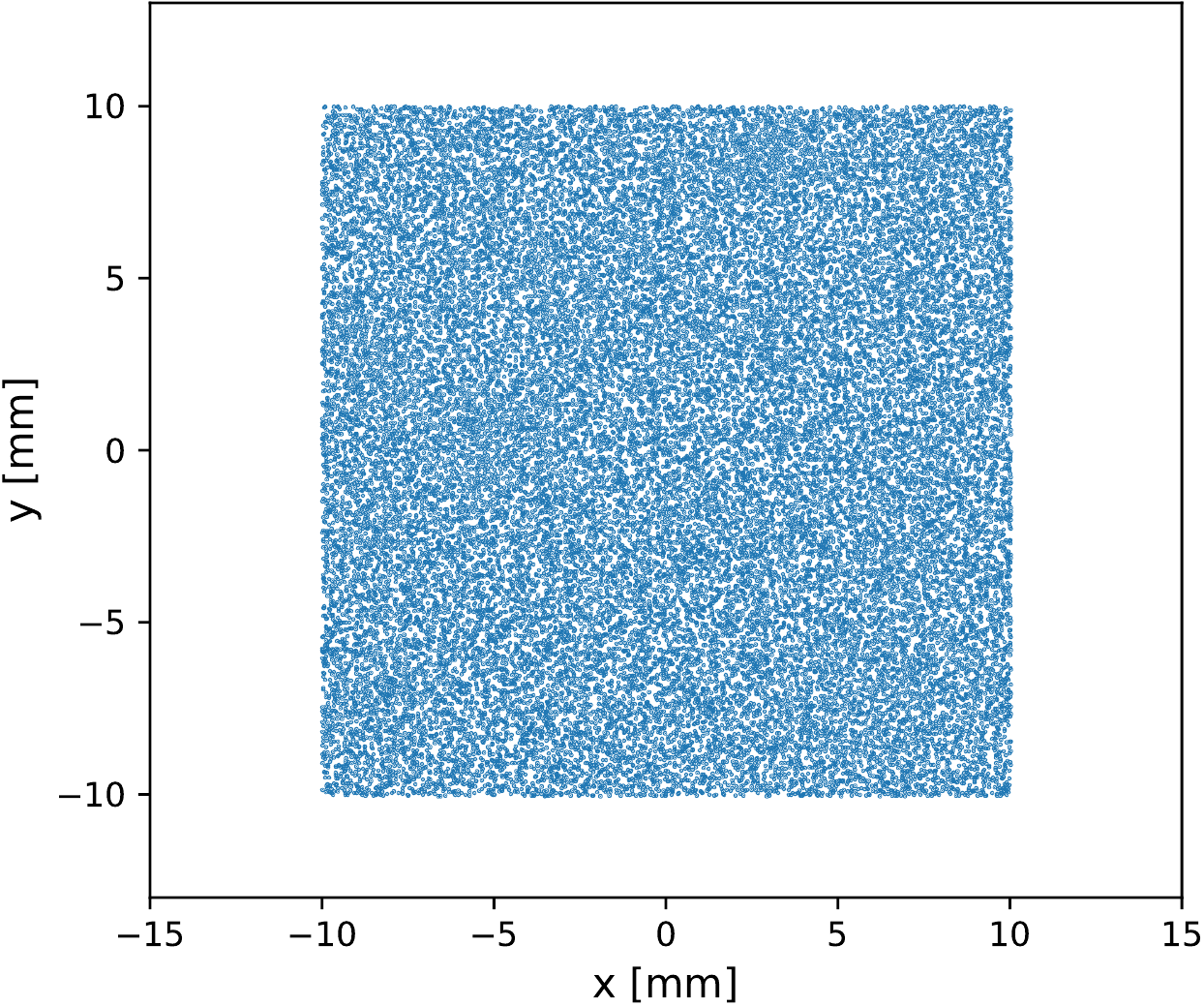}{0.46\textwidth}{}
          \fig{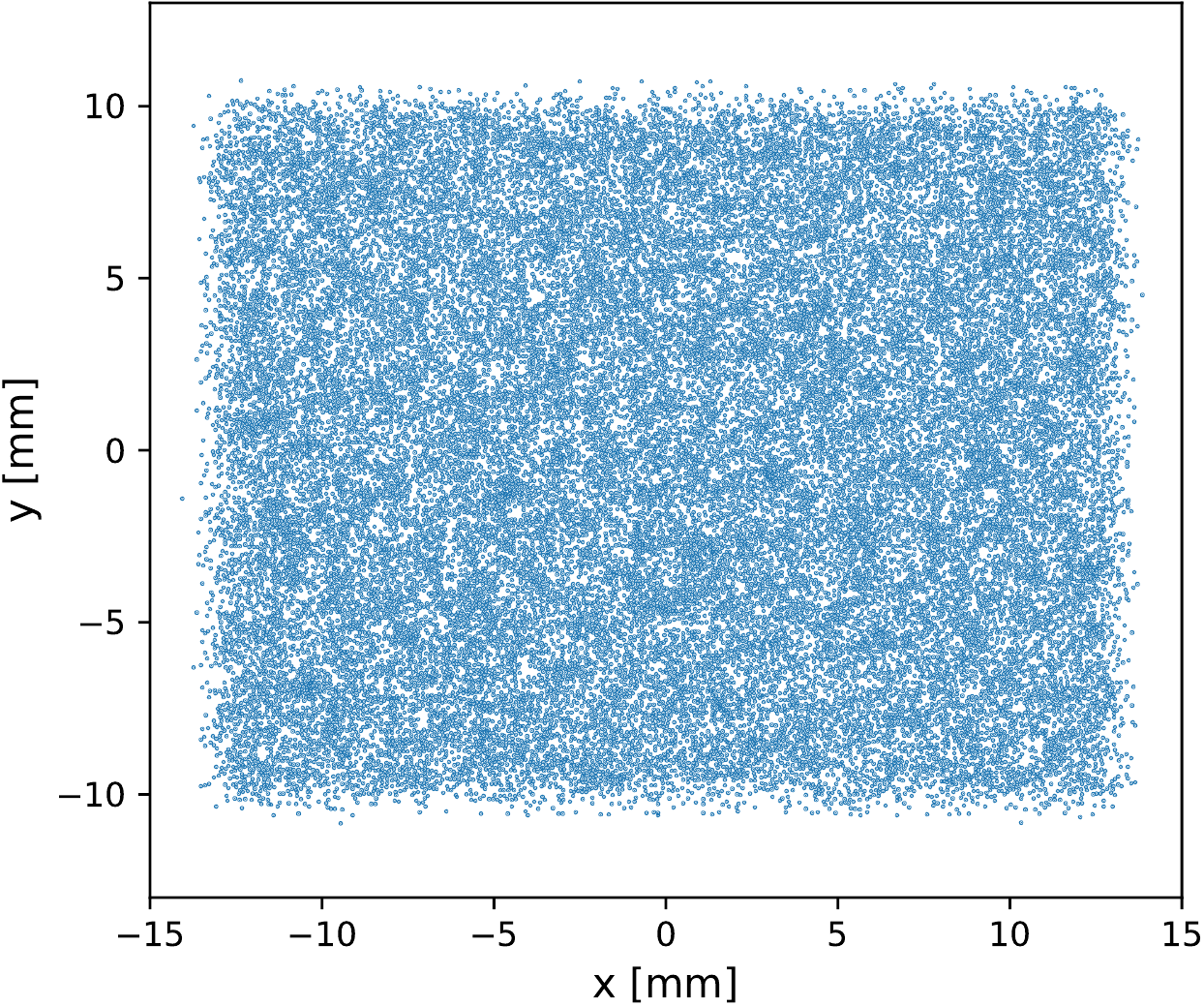}{0.46\textwidth}{}
          }
\gridline{\fig{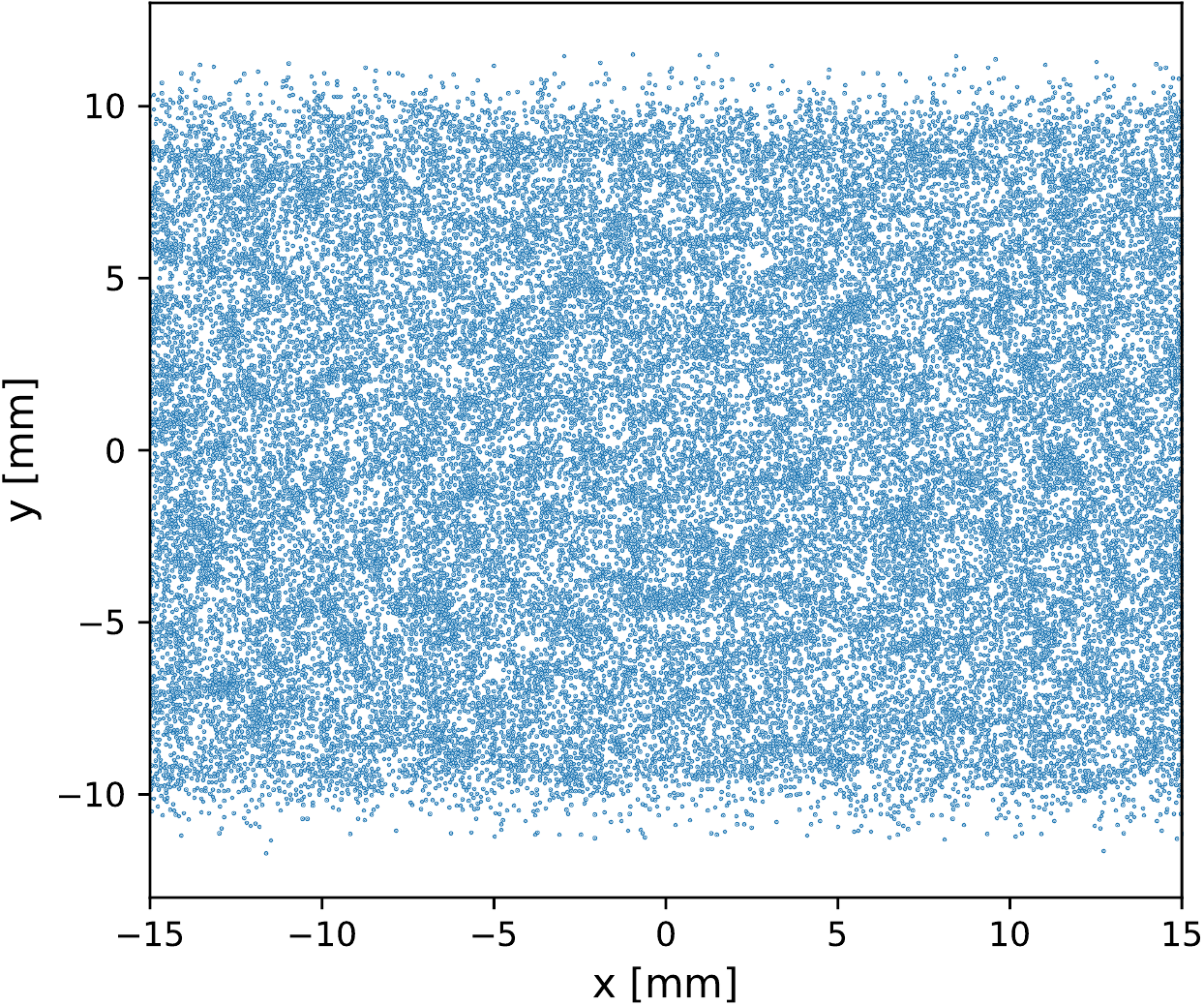}{0.46\textwidth}{}
          \fig{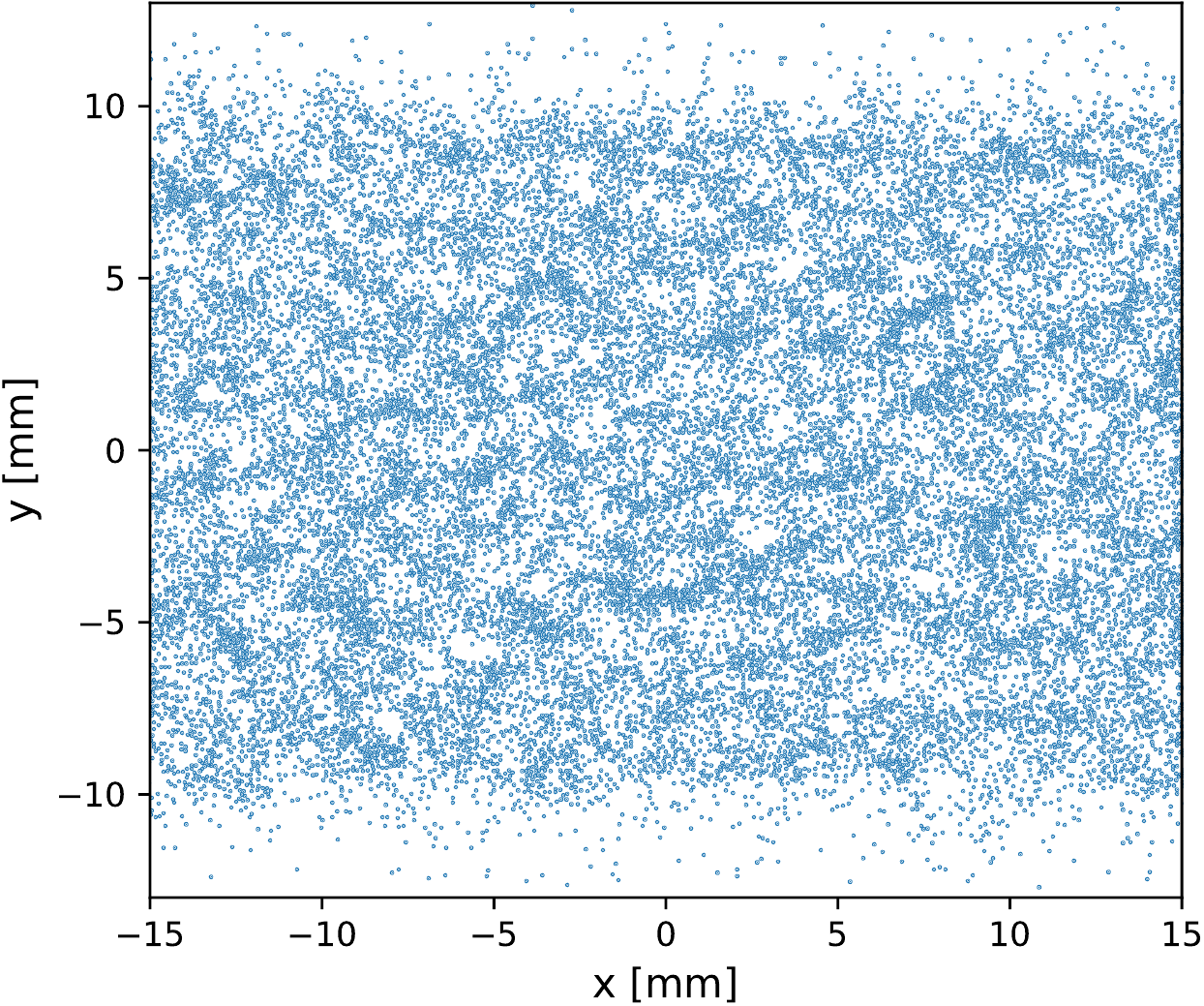}{0.46\textwidth}{}
          }
\gridline{\fig{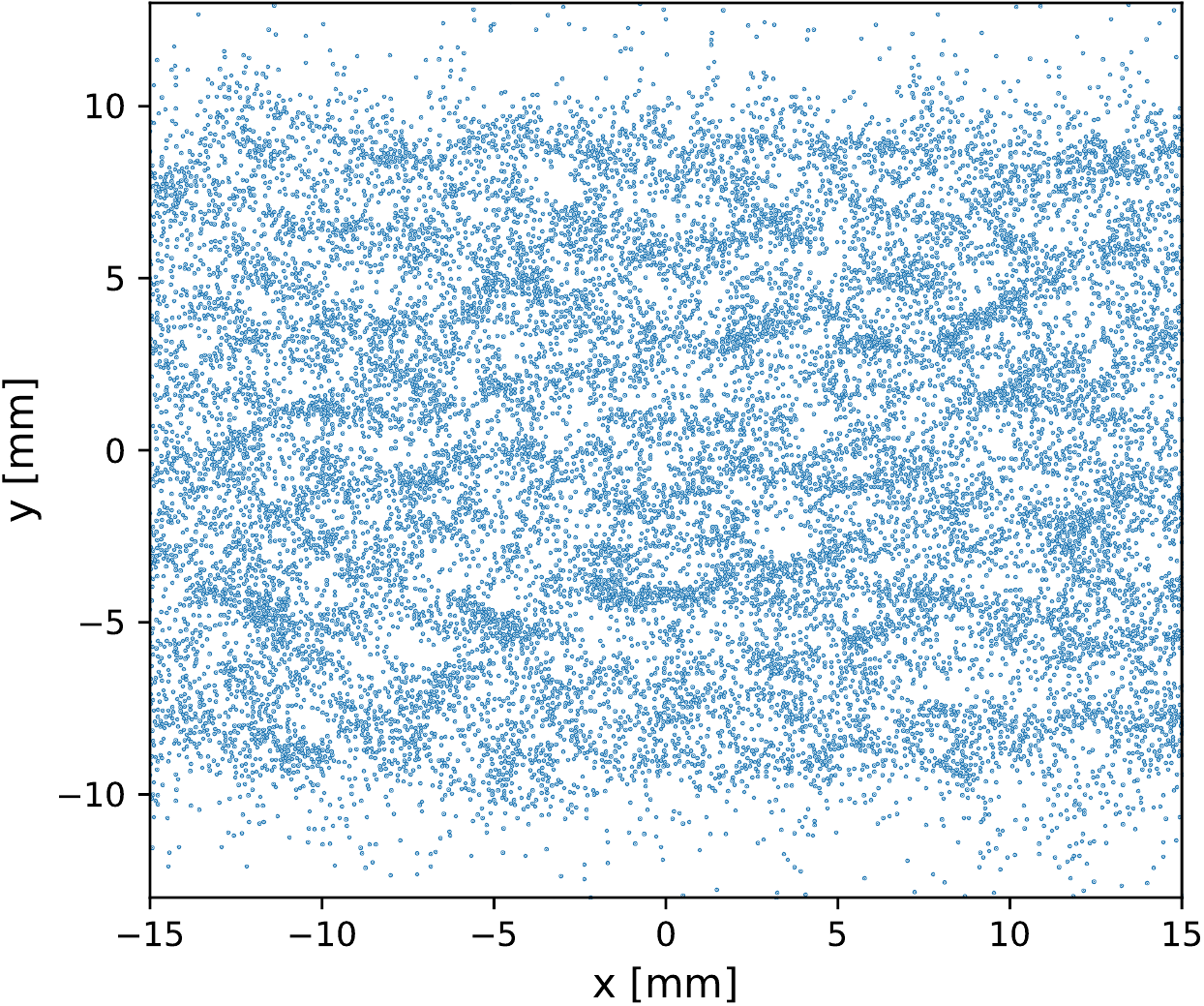}{0.46\textwidth}{}
          \fig{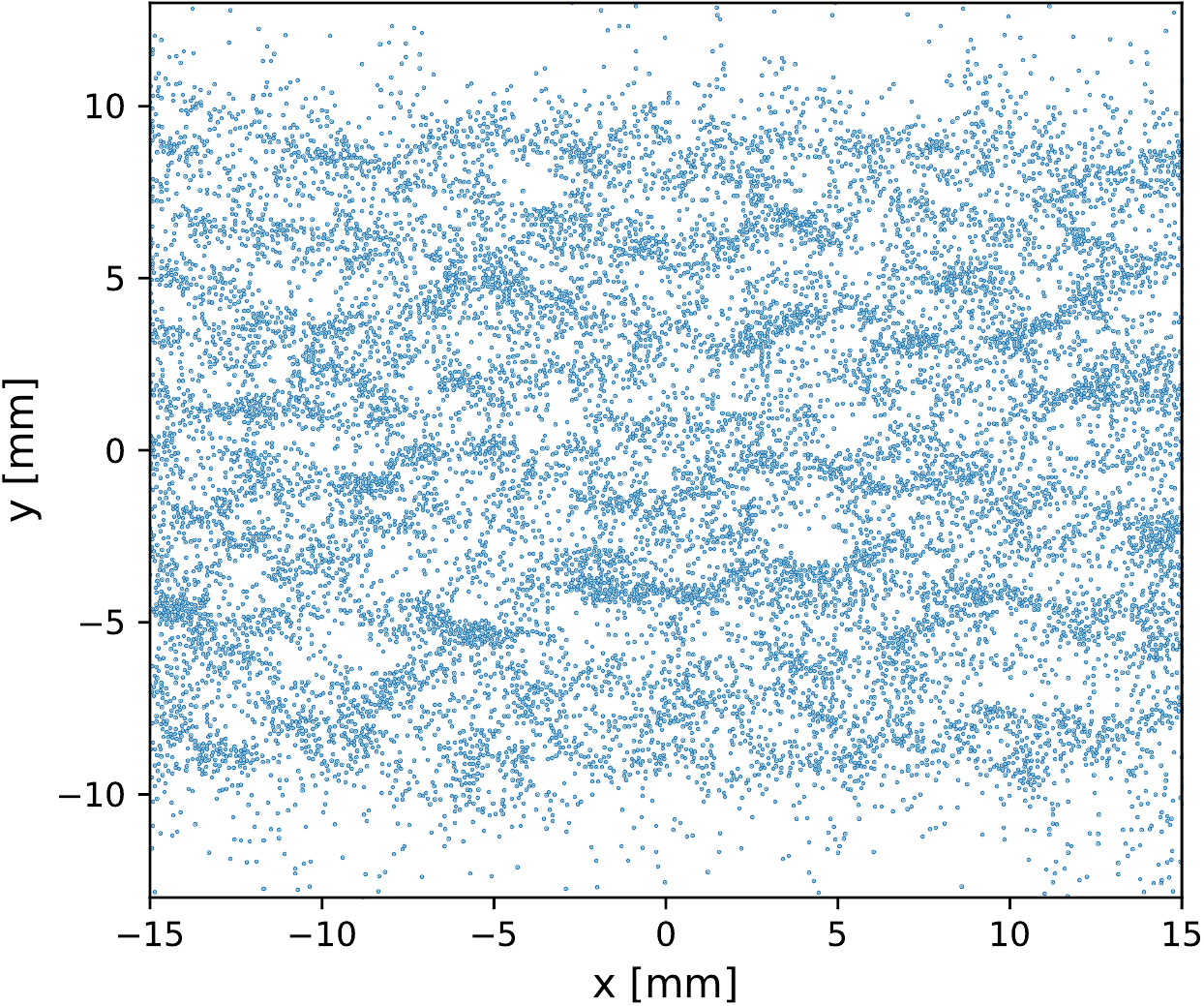}{0.46\textwidth}{}
          }
\caption{Snapshots of the particle distribution at times $t = $
0, 15, 30, 60, 90, and 120 $\rm{\mu s}$ (top left, top right, middle left, middle right, bottom left, and bottom right panels, respectively)
obtained from a simulation run (see Table~1 for the adopted parameter set)}.
\label{fig:simulation}
\end{figure*}

Figure~\ref{fig:simulation} shows snapshots of the particle distribution obtained from a single simulation run. Particle clustering is visible already at $t=15~\rm{\mu s}$ (upper right panel). At $t > 30~\rm{\mu s}$, the cluster pattern simply expands in the $x$ direction rather than growing further (middle right, bottom left, and bottom right panels), consistent with the result of our laboratory experiment. The pattern is mesh-like, demonstrating that inelastic collisions between particles at early times induce the formation of the mesh pattern in ejecta curtains observed in laboratory experiments including ours (see Section~\ref{sec:lab}). Because of the unidirectional expansion, the particle clusters constituting the mesh pattern are elongated along the $x$ axis. If the particles had higher expansion velocities in the $y$ direction, the resulting mesh pattern would be more elongated along the $y$ axis.

\begin{figure}
\gridline{\fig{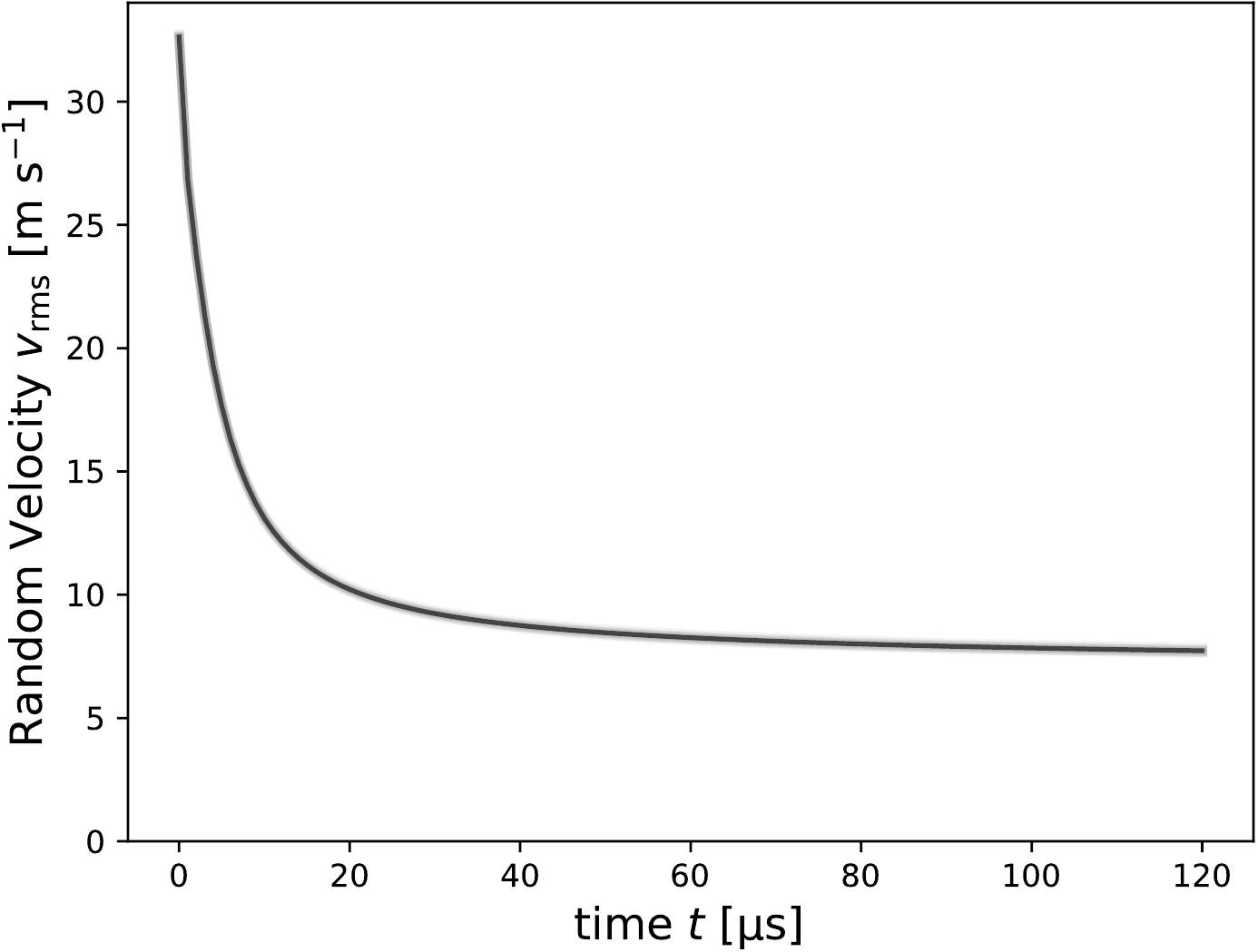}{0.45\textwidth}{}
          }
\caption{Time evolution of the RMS particle velocity $v_{\rm{RMS}}$ from the simulations. The thin gray lines are from 10 individual runs, while the thick black line shows the average over the 10 runs.}
\label{fig:sim_vel}
\end{figure}
Figure \ref{fig:sim_vel} shows the RMS random velocity $v_{\rm{RMS}}$ (Equation \ref{v_rms_sim}) averaged over the 10 runs as a function of $t$. We find that $v_{\rm RMS}$ decreases at $t < 20~ \rm{\mu s}$ when the mesh pattern is growing, and approaches to a constant at later times when the pattern is expanding. 
This serves as another piece of evidence for particles' inelastic collisions being the cause of pattern formation.

\begin{figure}
\gridline{\fig{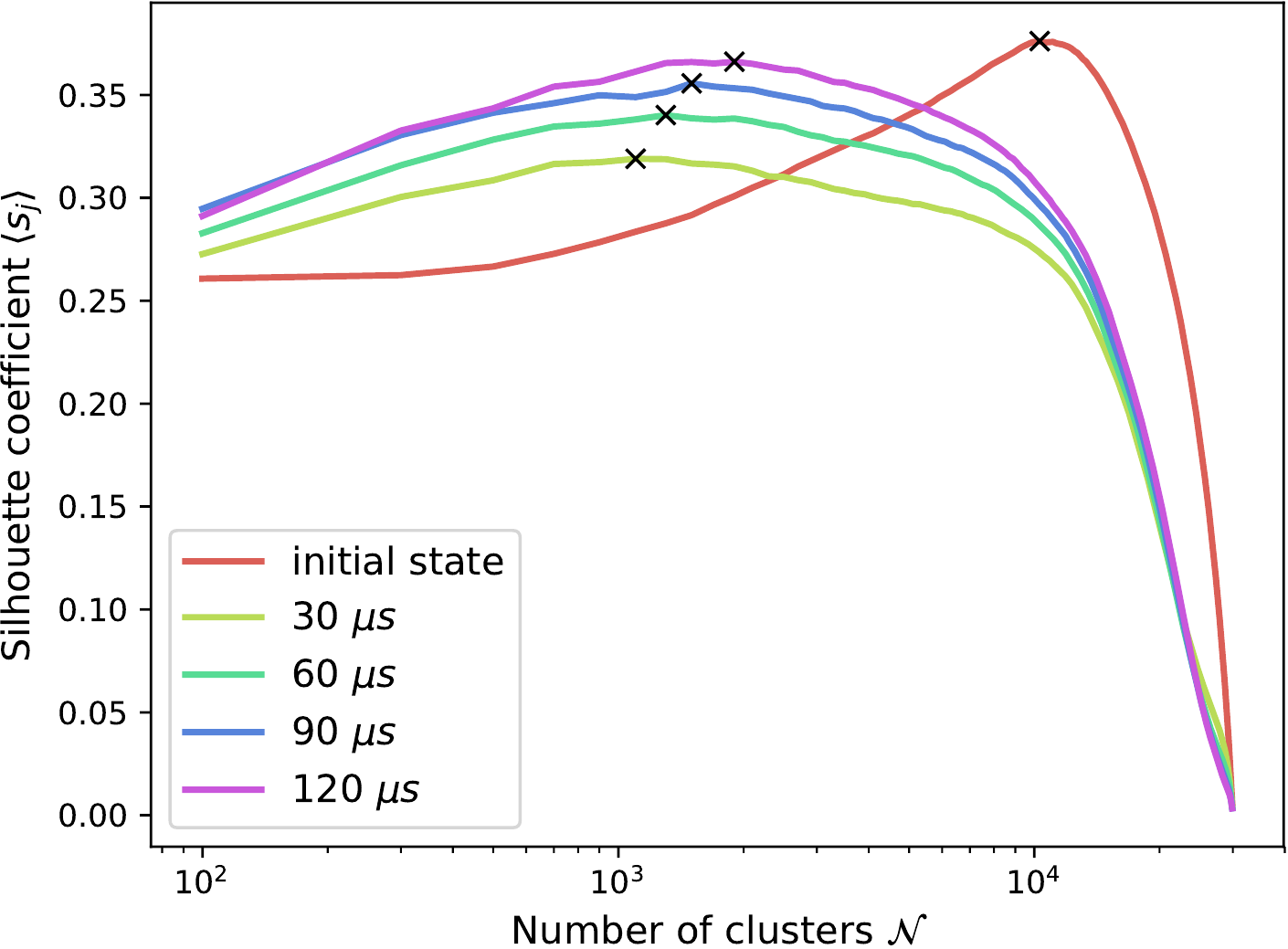}{0.45\textwidth}{}
          }
\caption{Particle-averaged silhouette coefficient $\langle s_j \rangle$ as a function of the number of clusters ${\cal N}$ at different times from a single simulation run. The cross symbols indicate the optimal number of clusters, at which $\langle s_j \rangle$ is maximized.}
\label{fig:silhouette}
\end{figure}
\begin{figure*}
\gridline{\fig{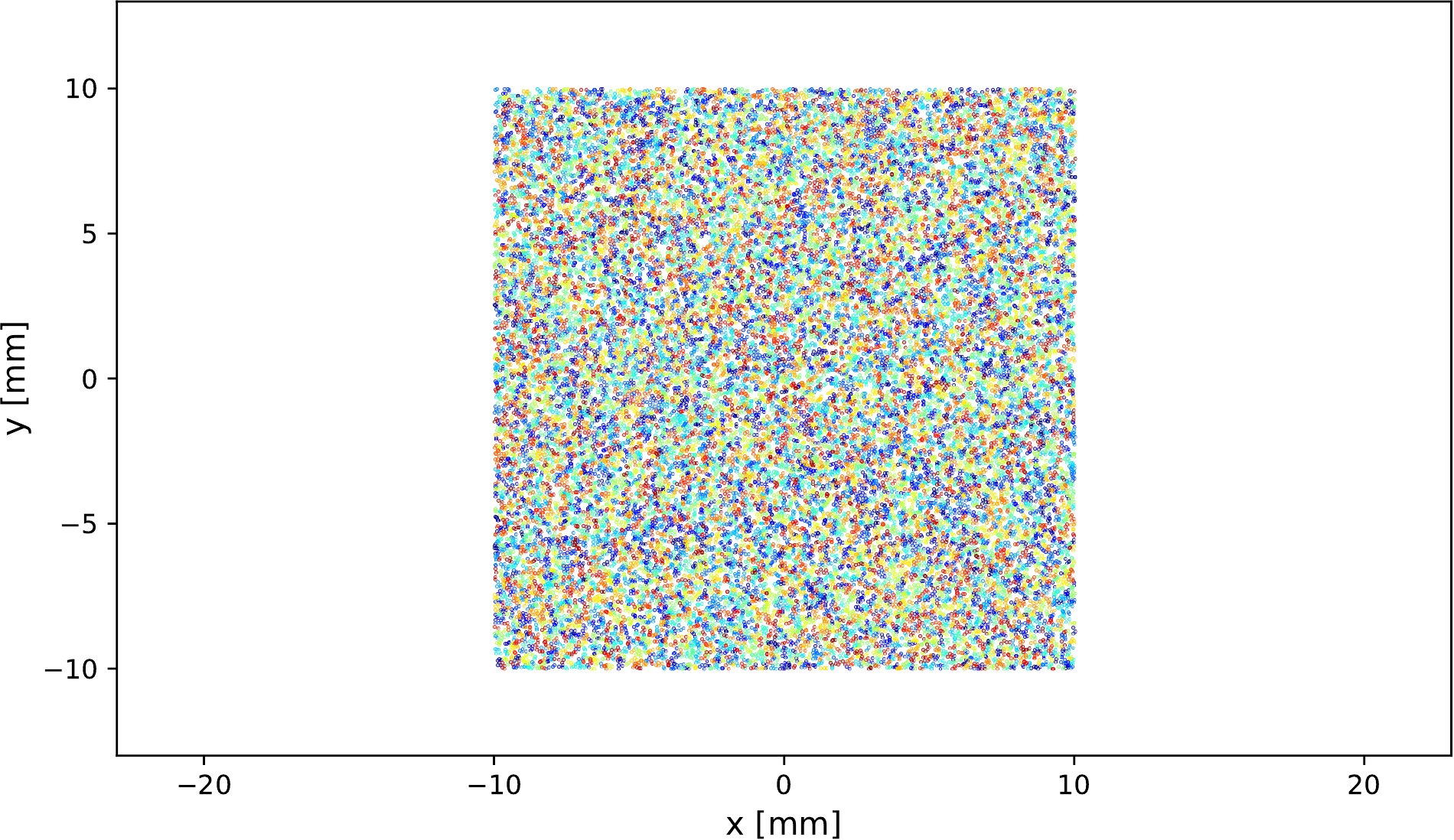}{1.0\textwidth}{(a)}
          }
\gridline{\fig{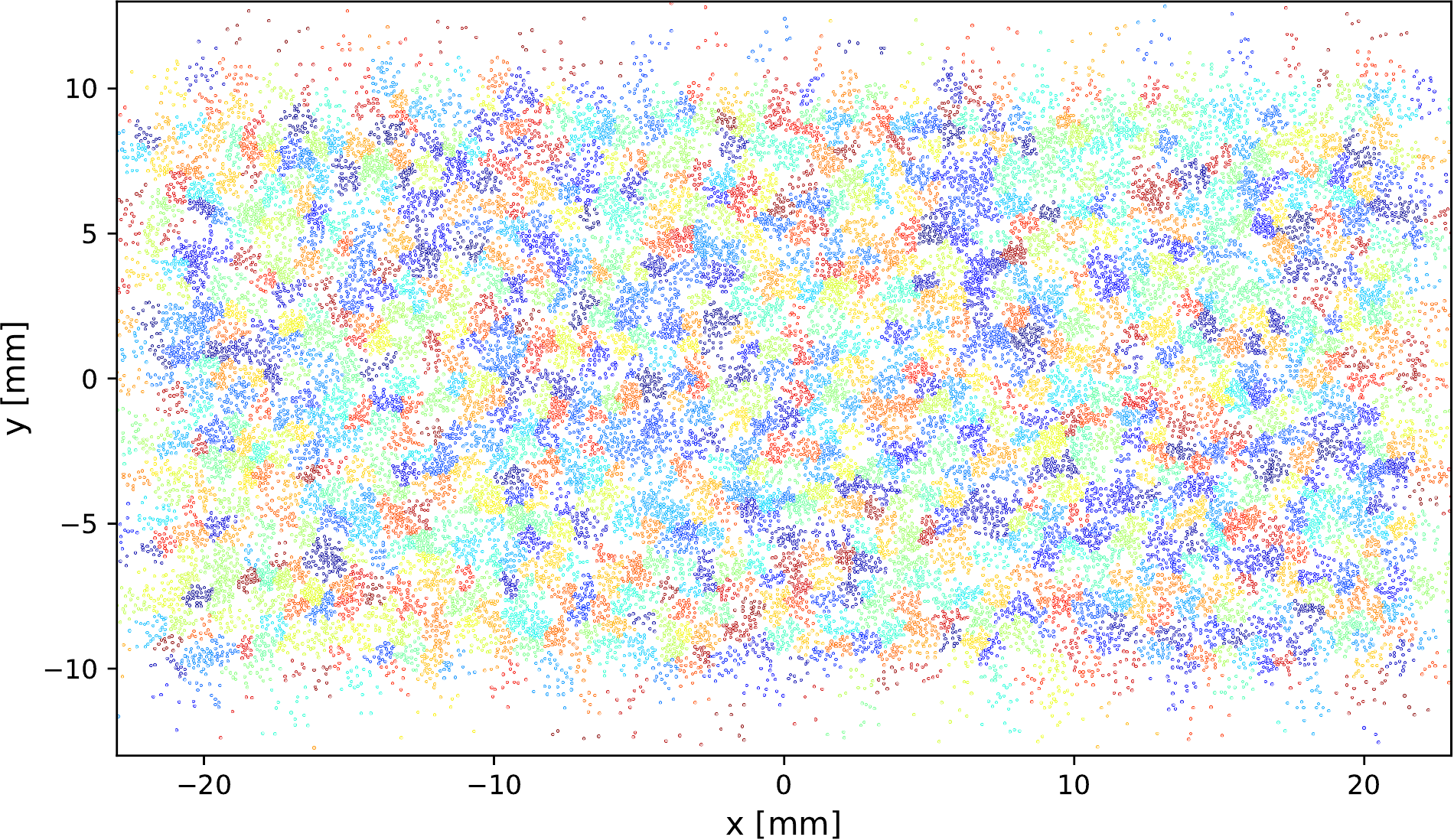}{1.0\textwidth}{(b)}
          }
\caption{Snapshots showing the particle cluster distribution at $t = 0$ and $60~\rm \mu s$ (panels (a) and (b), respectively) obtained by applying the cluster analysis to the particle distributions shown in the top left and middle right panels of Figure 3. 
The number of clusters ${\cal N}$  at $t = 0$ and $60~\rm \mu s$ are $10285$ and $1195$, respectively.
}
\label{fig:scatter}
\end{figure*}

To illustrate how we determine the optimal number of clusters from the cluster analysis, we plot in Figure~\ref{fig:silhouette} the particle-averaged silhouette coefficient $\langle s_j \rangle$ as a function of the number of clusters $\mathcal N$ at different times for a single run. Here, the values of $\langle s_j \rangle$ are calculated from $\mathcal N$ = 1 to 30000 for every 100 increment in $\mathcal N$. 
At all times, $\langle s_j \rangle({\cal N})$ has a single maximum, allowing us to uniquely define the optimal number of clusters as the number of clusters ${\mathcal N} (\langle s_j \rangle_{\rm {max}})$ that gives the maximum silhouette coefficient.
Figure~\ref{fig:scatter} illustrates how our cluster analysis works for the particular example shown in Section~\ref{fig:simulation} at $t = 0$ and 60~$\rm \mu s$.

\begin{deluxetable}{cccc}
\label{tb:cluseranalysis}
\tablecaption{Properties of the Optimal Clusters at Different Times $t$ (Averaged Over 10 Runs)}
\tablecolumns{4}
\tablewidth{1.0\columnwidth} 
\tablehead{
\colhead{$t~(\rm \mu s)$} &
\colhead{${\mathcal N} (\langle s_j \rangle_{\rm {max}})$} &
\colhead{$\overline{N}$} &
\colhead{$\delta N$}
}
\startdata
    0 & 10285 & 2.9 & 1.0 \\
    30 & 1265 & 23.7 & 8.7 \\
    60 & 1195 & 25.1 & 10.8\\
    90 & 1575 & 19.0 & 8.9\\
    120 & 1800 & 16.7 & 8.3 \\
\enddata
\end{deluxetable}
\begin{figure}
\gridline{\fig{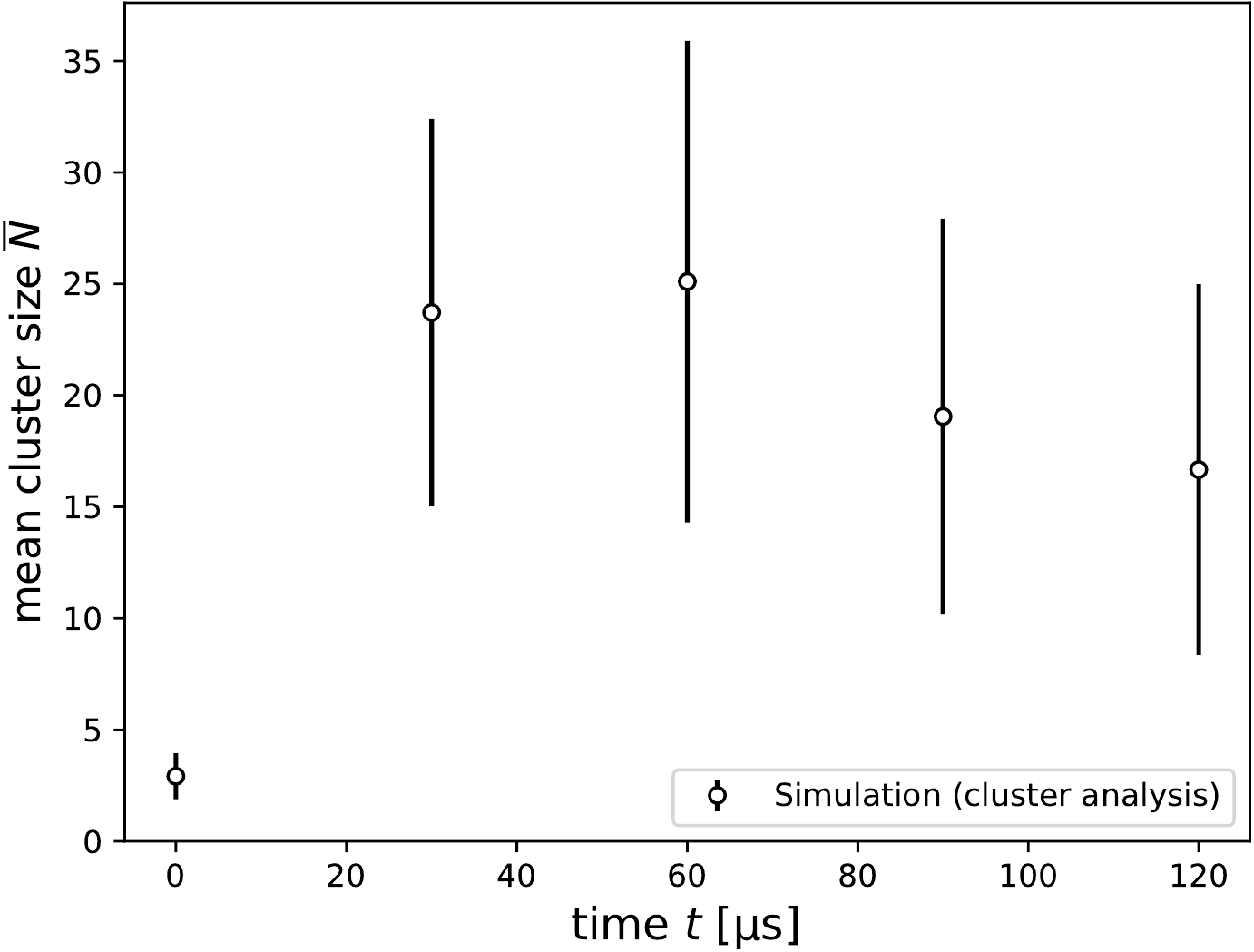}{0.45\textwidth}{}
          }
\caption{
Mean and standard deviation of the cluster size at different times $t$ (circles and error bars, respectively) obtained from the cluster analysis of the N-body simulations.}
\label{fig:Mass_sim}
\end{figure}

Table~\ref{tb:cluseranalysis} lists the optimal cluster number ${\mathcal N} (\langle s_j \rangle_{\rm {max}})$ averaged over 10 simulation runs, the corresponding mean cluster size $\overline{N} = N_{\rm tot}/{\cal N}(\langle s_j \rangle_{\rm {max}})$, and the standard deviation $\delta N$  of the cluster size averaged over the 10 runs. Here, we search for ${\mathcal N} (\langle s_j \rangle_{\rm {max}})$ for individual runs in two steps, first obtaining a rough estimate of $\langle s_j \rangle_{\rm max}$ by computing $\langle s_j \rangle$ from $\mathcal N$ = 1 to 30000 for every 100 increment in $\mathcal N$, and second obtaining a more precise value of $\langle s_j \rangle_{\rm max}$ by computing $\langle s_j \rangle$ for every 5 increment in $\mathcal N$ in the vicinity of ${\mathcal N} (\langle s_j \rangle_{\rm {max}})$  from the first step.
Figure~\ref{fig:Mass_sim} plots $\overline{N}$ and $\delta N$ versus $t$ given in Table~\ref{tb:cluseranalysis}. 
One can see that $\overline{N}$ increases in the first $30~\rm \mu s$, when the RMS particle random velocity decreases (see Figure~\ref{fig:sim_vel}), indicating that our cluster analysis well captures pattern formation through inelastic collisions.
At later times, the cluster mass plateaus out, reflecting the collisionless expansion of the system. Close inspection shows that the derived cluster mass gradually decreases after $t=60~\rm{\mu s}$. We interpret this as suggesting that our cluster analysis does not fully resolve the clusters at $t \la 60~\rm \mu s$.

\begin{figure*}
\gridline{\fig{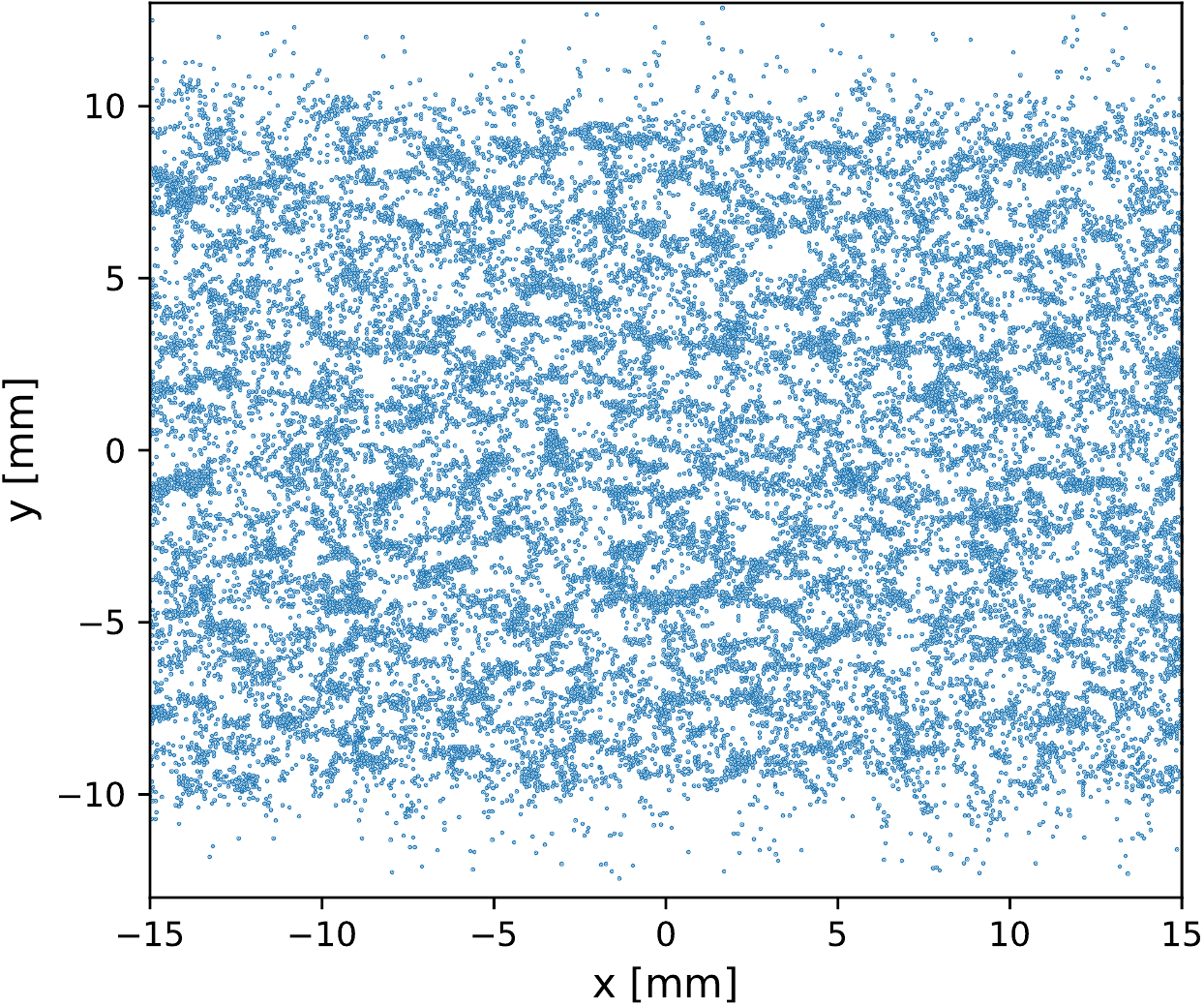}{0.46\textwidth}{}
          \fig{Article_fig_sim_last_60us-crop.pdf}{0.46\textwidth}{}
          }
\gridline{\fig{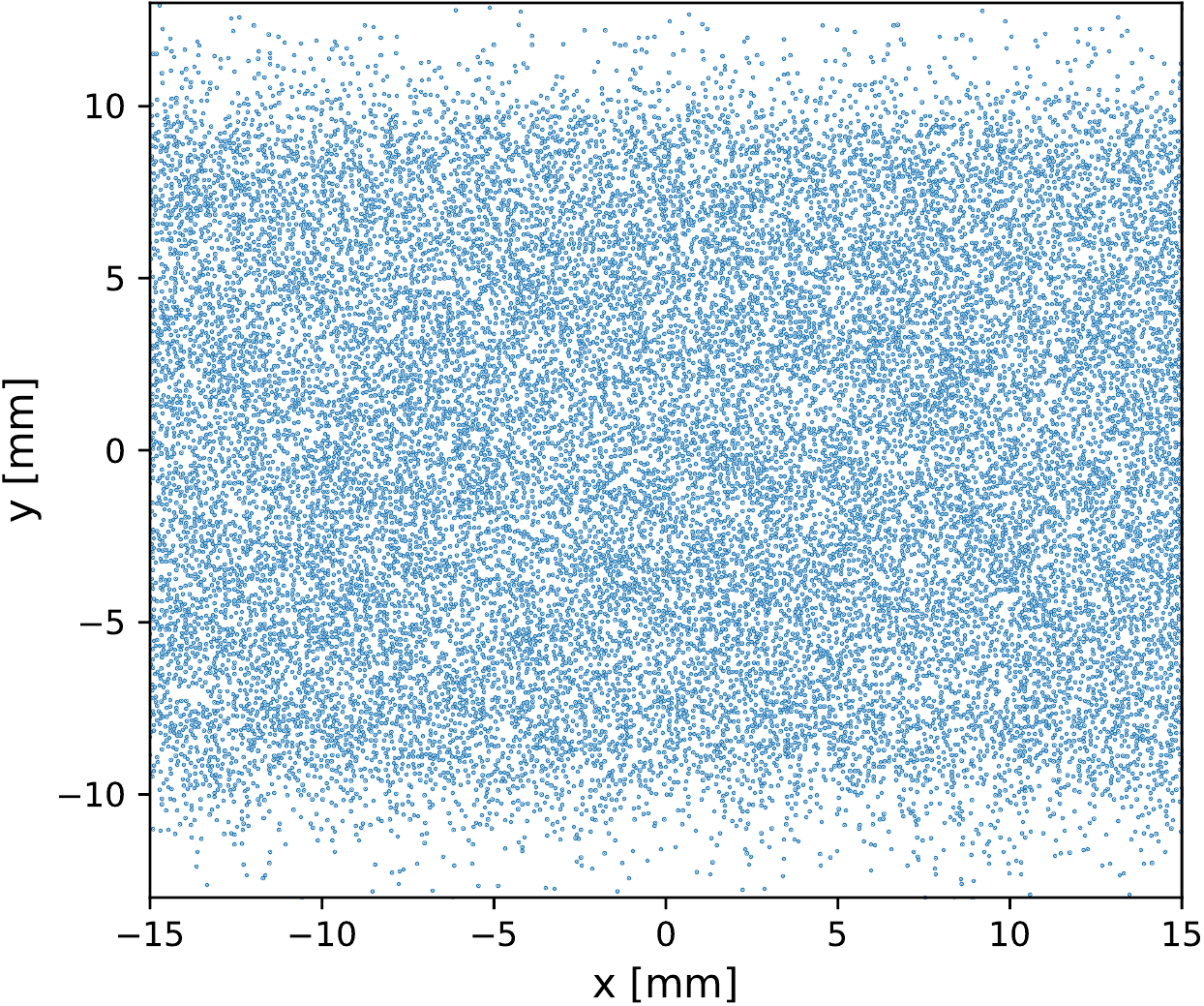}{0.46\textwidth}{}
          \fig{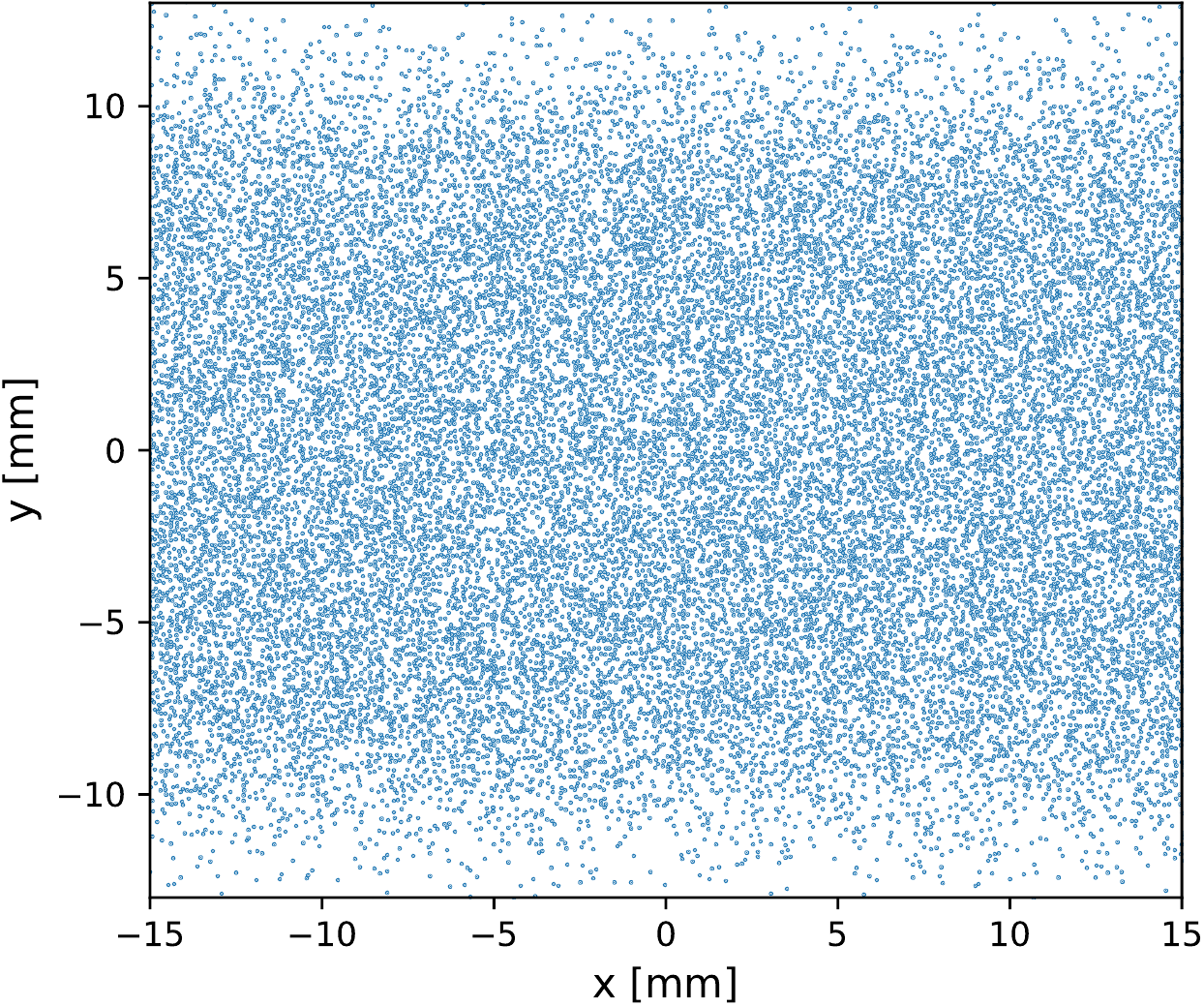}{0.46\textwidth}{}
          }
\caption{Snapshots of the particle distribution at a time $t = $
60 $\rm{\mu s}$ from simulation run with $e = $ 
0.1, 0.6, 0.9 and 1.0 (top left, top right, bottom left, and bottom right panels, respectively). The parameters other than $e$ are fixed as listed in Table \ref{tb:parameter}.}
\label{fig:simulation_change_e}
\end{figure*}

Figure \ref{fig:simulation_change_e} compares the particle distribution at $t = 60 ~{\rm \mu s}$ from simulations of different values of $e$.
We see that particle clusters become more appreciable as $e$ decreases, again supporting the idea that the pattern formation is driven by inelastic collisions. The observed trend is also qualitatively consistent with the simulation results by \citet[their Figure 6]{KADONO2015215}.

\section{Analytic Model}
\label{sec:modeling}

Our simulations presented in the previous section confirm the expectation by \citet{KADONO2015215,KADONO2020113590} that the mesh pattern in ejecta evolves in two stages, initial growth through inelastic particle collisions and subsequent geometric expansion with no particle collision.  
However, it is yet to be addressed when the initial growth stage is terminated and what initial conditions set the final size of the clusters constituting the pattern. 
To address these questions, we here construct an analytic model for the growth of particle clusters in an expanding particle system.

\subsection{Model Description}
\label{subsec:modeling_equ}

As discussed in the previous section, it is clear that  
inelastic particle collisions are the cause of the cluster formation seen in our simulations. Because particles constituting clusters experience inelastically collide with each other frequently, we can expect that colliding clusters lose their initial collision energy quickly.  
Based on this consideration, our analytic model assumes that clusters coalesce perfectly whenever they collide.
As discussed in Section~\ref{sec:discussion}, this assumption is valid if the coefficient of restitution of the particles is sufficiently low.

For the moment, we also assume the random velocity component dominates over the non-isotropic velocity component arising from the system's expansion motion. This assumption breaks down at late times where the expansion motion terminates the collisional evolution of the system. We account for this effect at the end of this subsection. 
To make our model as simple as possible, we approximate all clusters as equal-sized spheres and only consider their one-dimensional motion. 

\subsubsection{Relationship Between the Cluster Velocity Dispersion and Mass}
The assumptions made above gives a simple relationship between the cluster velocity dispersion and cluster mass. 
Let the velocities of two colliding clusters 1 and 2 be $V_1$ and $V_2$ and the velocity of the merged cluster be $V'$. 
Using the conservation of momentum law and the assumption that the clusters are approximately equal in size, we have 
\begin{equation}\label{merge}
  V'=\frac{1}{2}(V_1+V_2).
\end{equation}
The assumption that the isotropic, random motion dominates the clusters' velocities gives  
\begin{equation}\label{v_assumption}
  \begin{cases}
    \overline{V_1} = \overline{V_2}  = 0 ~,\\
    \overline{V_1^2} = \overline{V_2^2}  = \sigma^2 ~, \\
    \overline{V_1 V_2} = 0,
  \end{cases}
\end{equation}
where $\sigma$ is the velocity dispersion of the clusters and the overlines denote ensemble averages over all clusters.
Using Equations~(\ref{merge}) and~(\ref{v_assumption}), the velocity dispersion of the clusters after a single collision,
$\sigma' \equiv (\overline{V'^2})^{1/2}$, can be written as 
\begin{equation}\label{rms_after_col}
  \sigma'= \frac{1}{\sqrt{2}}\sigma ~.
\end{equation}
Equation~(\ref{rms_after_col}) shows that the velocity dispersion of the clusters decreases by a factor of $1/\sqrt{2}$ after every merging collision.
Because the cluster mass increases by a factor of 2 after every collision, one can relate $\sigma$ to the cluster mass as
\begin{equation}
  \label{Model_Vel}
  \sigma(\overline{N})= \sigma_0\left(
  \frac{N_0}{\overline{N}}\right)^{1/2},
\end{equation}
where $\overline{N}$ is the average number of particles per cluster (which is equal to the mean cluster mass in units of the monomer mass) and  $\sigma_0$ and $N_0$ are the initial values of $\sigma$ and $\overline{N}$, respectively.

\subsubsection{Time Evolution of the Cluster Mass}
The time evolution of $\overline{N}$ through cluster mergers can be described by 
\begin{equation}\label{diff_N1}
    \frac{d\overline{N}}{dt}=\frac{\overline{N}}{\tau} ~,
\end{equation}
where $\tau$ is the mean collision time.
For spherical clusters moving on a two-dimensional space, $\tau$ can be estimate as
\begin{equation}\label{tau}
    \tau = \frac{1}{4\overline{R}n\sigma},
\end{equation}
where $\overline{R}$ and $n$ are the average radius and surface number density of the clusters (the factor $4\overline{R}$ represents the collisional cross section for equal-sized disks on a plane).
Assuming that each two-dimensional cluster is densely packed, $\overline{R}$ is related to $\overline{N}$ and the particle radius $r$ as $\overline{N} = \pi \overline{R}^2/(\pi r^2)$, or
\begin{equation}\label{R_def}
    \overline{R} = r\overline{N}^{1/2}.
\end{equation}

Since the number of clusters in the system is $\approx N_{\rm tot}/\overline{N}$, the cluster surface number density $n$ can be estimated as 
\begin{equation}\label{R&n}
n = \frac{N_{\rm tot}}{L_V L_H \overline{N}},
\end{equation}
where $L_H$ and $L_V$ are the horizontal and vertical length of the system.
We assume that the vertical system size is approximately constant whereas the horizontal system size increases at constant expansion rate $\Omega$, 
\begin{equation}\label{expansion}
    L_H = L_{H_0}(1+\Omega t),
\end{equation}
with $L_{H_0}$ being the initial value of $L_H$. 
Substituting Equations~(\ref{Model_Vel}) and  \eqref{tau}--(\ref{expansion}) into Equation~\eqref{diff_N1}, we obtain
\begin{equation}
    \label{Model_Mass}
    \frac{d\overline{N}}{dt} = \frac{\dot{N}_0}{1+ \Omega t} ~,
\end{equation}
where $\dot{N}_0$ is a constant defined by
\begin{equation}\label{alpha}
    \dot{N}_0 = \frac{4r N_0^{1/2} N_{\rm tot} \sigma_0}{L_V L_{H_0}}~.
\end{equation}
The solution to eq.~(\ref{Model_Mass}) is
\begin{equation}\label{analytic_solution}
    \overline{N}(t) = N_0 + \frac{\dot{N}_0}{\Omega} \ln(1+\Omega t),
\end{equation}
where $N_0$ is the initial value of $\overline{N}$. 
The set of Equations (\ref{Model_Vel}) and (\ref{Model_Mass}) predicts $\sigma$ and $\overline{N}$ as a function of $t$ under the assumption that the expansion motion of the system has negligible effect on the cluster collision velocity. 

\subsubsection{``Freezeout'' of the Cluster Pattern}
\label{sec:freezeout}
For late times where the expansion motion of the system dominates over the random motion of individual clusters, we expect that the clusters can no longer grow collisionally. The cluster pattern then should ``freeze out,'' i.e., should simply expand self-similarly with time as observed in our experiment and simulation. 
We expect the freezeout to occur when the collisional timescale $\tau$ becomes longer than the expansion timescale $\sim \Omega^{-1}$. 
We thus define the freezeout time $t_{\rm freeze}$ by  
\begin{equation}\label{eq:tfreeze}
    \tau(t>t_{\rm freeze}) > \frac{1}{\Omega}~.
\end{equation}
To account for the freezeout, we stop the evolution of $\sigma$ and $\overline{N}$ stops at $t = t_{\rm freeze}$. 

The model confirms the expectation from our laboratory expriment that the cluster growth is completed on a timescale comparable to the expansion timescale $1/\Omega$.
By definition, the freezeout time satisfies $\tau(t=t_{\rm freeze}) = 1/\Omega$. Using Equations~\eqref{R_def}--\eqref{expansion}, we rewrite $\tau$ as
\begin{equation}
\label{eq:tau_sim}
\tau = 
\frac{L_V L_{H_0}(1 + \Omega t)}{4r N_{\rm{tot}} \sigma_0 N_0^{1/2}}
\overline{N}.
\end{equation}
For $\overline{N} \gg 1$, we have
\begin{equation}
\label{eq:Nbar_sim}
\overline{N} \approx \frac{\dot{N}_0}{\Omega}\ln(1+\Omega t),
\end{equation}
Substituting Equations \eqref{alpha} and \eqref{eq:Nbar_sim} into Equation \eqref{eq:tau_sim}, we obtain
\begin{equation}\label{tau_order}
    \tau \approx \frac{(1+\Omega t)\ln(1 + \Omega t)}{\Omega}.
\end{equation}
Therefore, the condition $\tau(t=t_{\rm freeze})=1/\Omega$ implies 
\begin{equation}\label{requirement_omega_t}
    (1 + \Omega t_{\rm{freeze}})\ln (1 + \Omega t_{\rm{freeze}}) \approx 1,
\end{equation}
from which we find
$t_{\rm{freeze}} \approx 0.76/{\Omega}$. We note that the freezeout time depends only on $\Omega$.

\subsubsection{Final Cluster Mass and Velocity Dispersion as a Function of Initial Conditions}
From Equation (\ref{analytic_solution}) and $t_{\rm{freeze}} \approx 0.76/{\Omega}$, we can write the final cluster size as a function of the initial cluster size and random velocity dispersion as
\begin{equation}\label{final_cluster_size}
    \overline{N}(t_{\rm freeze}) \approx N_0 + 0.57\frac{\dot{N}_0}{\Omega}.
\end{equation}
Substituting Equation (\ref{final_cluster_size}) into Equation (\ref{Model_Vel}), we obtain the final velocity dispersion of the clusters as
\begin{equation}
  \label{final_cluster_vrms}
  \sigma(t_{\mathrm{freeze}}) \approx \sigma_0\left(
  \frac{N_0}{N_0 + 0.57\dot{N}_0/\Omega}\right)^{1/2}.
\end{equation}

\subsection{Comparison with Simulations}
\label{subsec:modeling_comparison}

\begin{figure*}
\gridline{\fig{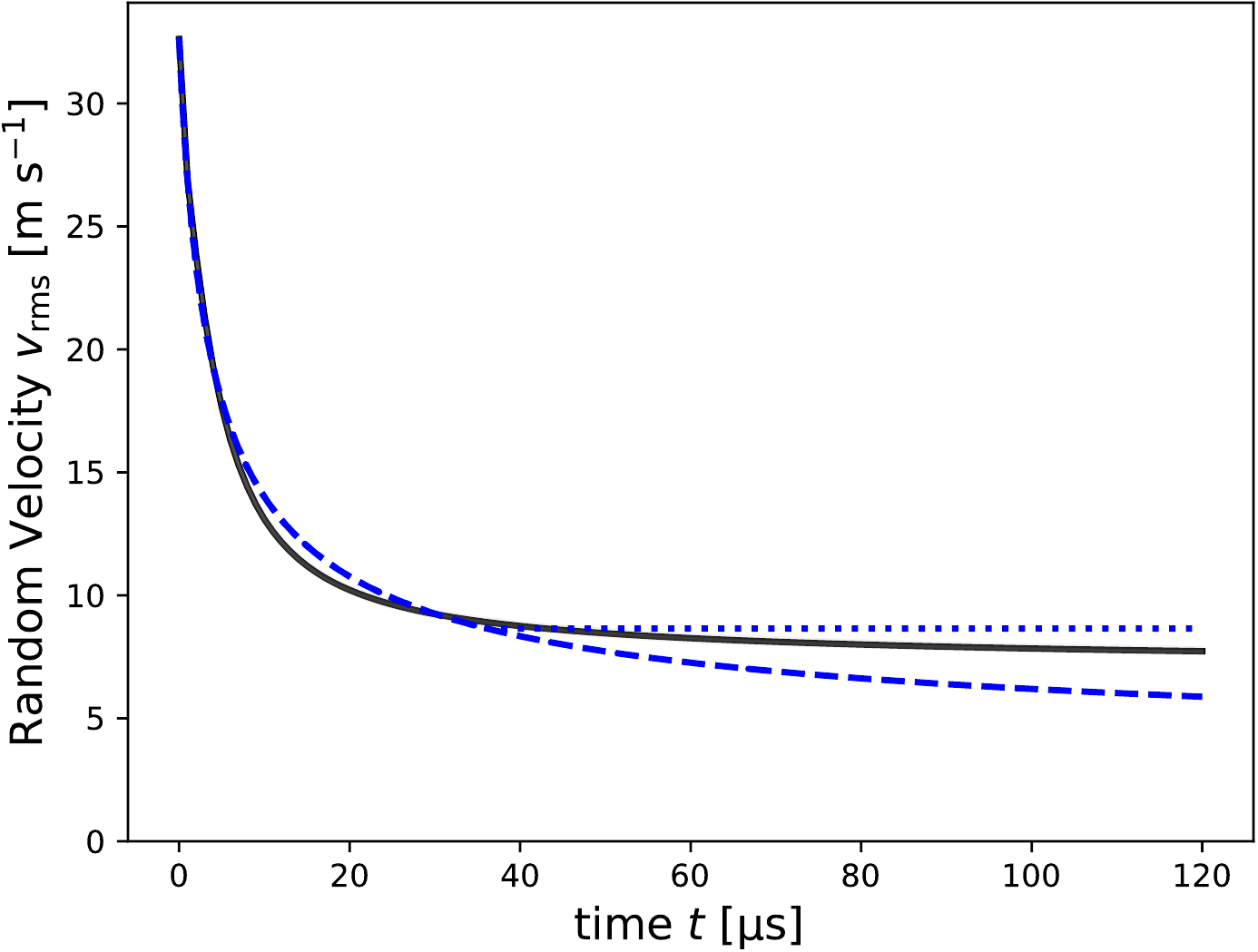}{0.45\textwidth}{}
          \fig{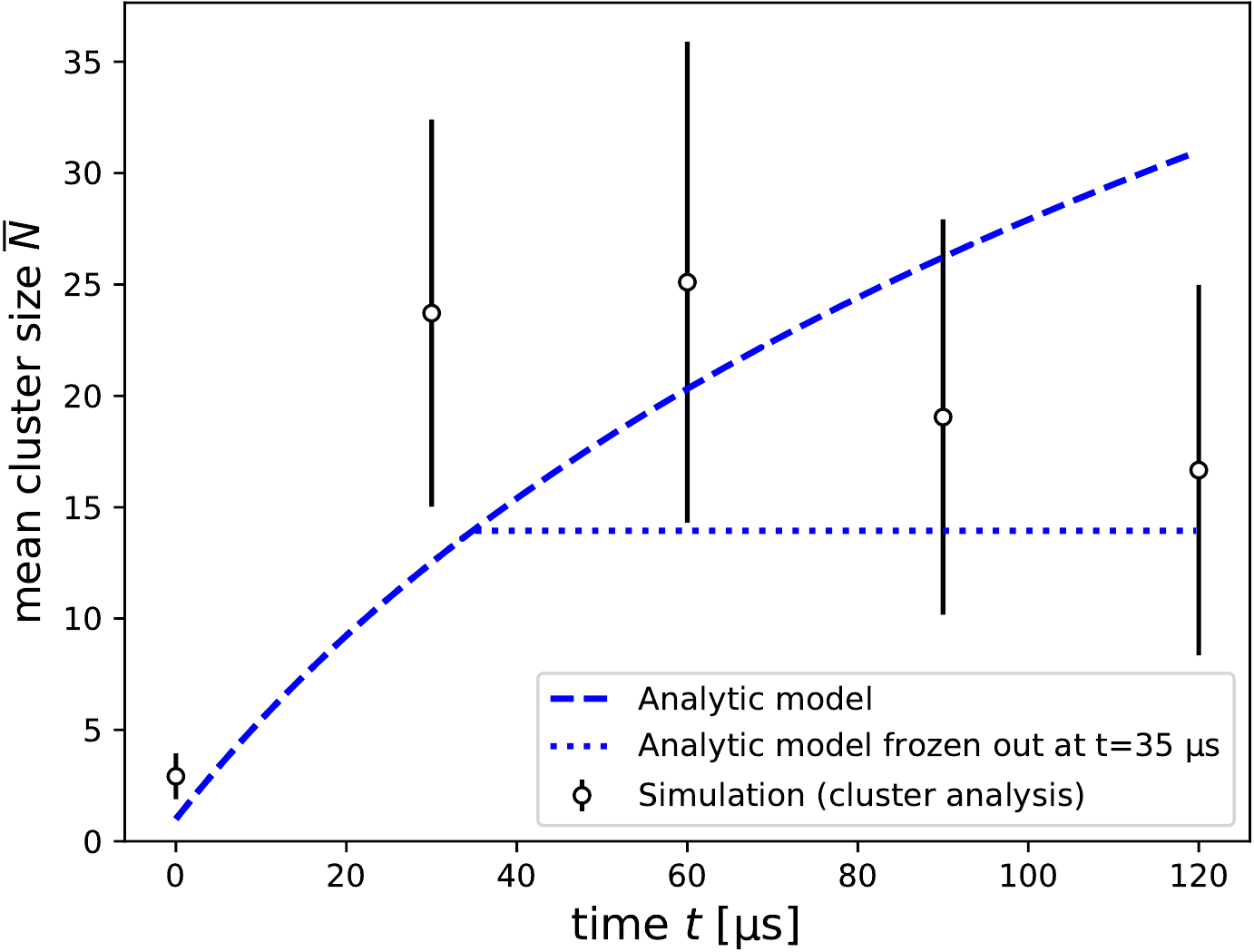}{0.45\textwidth}{}
          }
\caption{
Cluster random velocity (left panel) and mean cluster size (right panel) as a function of time $t$ from our simulation (solid lines; same as Figures~\ref{fig:sim_vel} and \ref{fig:Mass_sim}), compared with the prediction from our analytical model with and without the freezeout of the cluster growth at $t > 35~\rm{\mu s}$ (dashed and dotted lines, respectively). 
}
\label{fig:vel}
\end{figure*}

Now we test the analytic model derived above against our numerical simulations.
Figure~\ref{fig:vel} compare the cluster random velocity and mean cluster size measured in the simulations with those predicted from the analytic model (Equations~\eqref{Model_Vel} and \eqref{Model_Mass}, respectively) with and without the freezeout of the clusters at $t>t_{\rm freeze}$. 
The parameters in the model are set to $N_0 = 1$, $L_V = L_{H_0} = 20~\rm mm$, and $\sigma_0 = 32.6~{\rm m~s^{-1}}$. The value of $\sigma_0$ is taken to be close to the initial RMS speed of the particles in the simulations.
The assumed parameters give $\dot{N}_0 = 0.49~\rm{\mu s^{-1}}$.
The definition of the freezeout time (Equation~\eqref{eq:tfreeze}) gives $t_{\rm{freeze}} = 35~\rm{\mu s}$ for these parameter choices.

We find that the analytic model neglecting the freezeout well reproduces the evolution of $v_{\rm RMS}$ at least at early times $t \la 30~\rm \mu s$. 
This indicates that energy dissipation of particle clusters through perfect coalescence explains the decrease of the random velocity.
At $t \ga 30~\rm \mu s$, the random velocity from the simulation decreases more slowly than predicted from the model, suggesting that this model overestimates the frequency of cluster collisions at late times.

The analytic model neglecting the freezeout reproduces the mean cluster size in the simulations to within a factor of 2.
The agreement is less good than for the random velocity, in particular at $t \le 60~\rm{\mu s}$, where $\overline{N}$ from the simulation is considerably higher than predicted by the model.
This is likely because the clusters are so close that our cluster analysis using the silhouette coefficient $s_j$ (Equation~\eqref{eq:silhouette}) does not fully resolve them.
Density-based cluster analysis algorithms could mitigate this issue, but would introduce additional free parameters (for example, the DBSCAN algorithm \citep{10.5555/3001460.3001507} requires the detection radius and the minimum number of cluster members).
We defer further investigation of this issue to future work.
At $t \ga 80~\rm \mu s$, the model predicts higher $\overline{N}$ than observed in the simulation. 
This is another indication that cluster collisions at late times are less frequent than expected by the model.

Accounting for the freezeout of the clusters at $t > t_{\rm freeze}$ significantly improves the model predictions at late times (see the dotted lines in Figure~\ref{fig:vel}).
Substituting the initial conditions of the simulation (Table \ref{tb:parameter}, $\sigma_0 = 32.6~\rm{m~s^{-1}}$) into Equations (\ref{final_cluster_size}) and (\ref{final_cluster_vrms}), we obtain $\bar{N}(t_{\mathrm{freeze}}) = 14.9$, ~$\sigma(t_{\mathrm{freeze}}) = 8.4~\mathrm{m~s^{-1}}$. These model predictions are consistent with the final values from the simulation.

To summarize this section, we have shown that the initial growth of particle clusters in an expanding particle system terminates when the collisional timescale exceeds the expansion timescale $1/\Omega$. This occurs at $t_{\rm freeze} \sim 1/\Omega$ after the clusters start growing. Our analytic model quantitatively reproduces the evolution of the cluster mass and velocity dispersion as a function of the initial conditions.

\section{Discussion}
\label{sec:discussion}

\subsection{Validity and Limitations of the Model}
\label{subsec:discussion_model}
In this section, we discuss the validity and limitations of our analytic model presented in Section~\ref{sec:modeling}.

\begin{figure}
\gridline{\fig{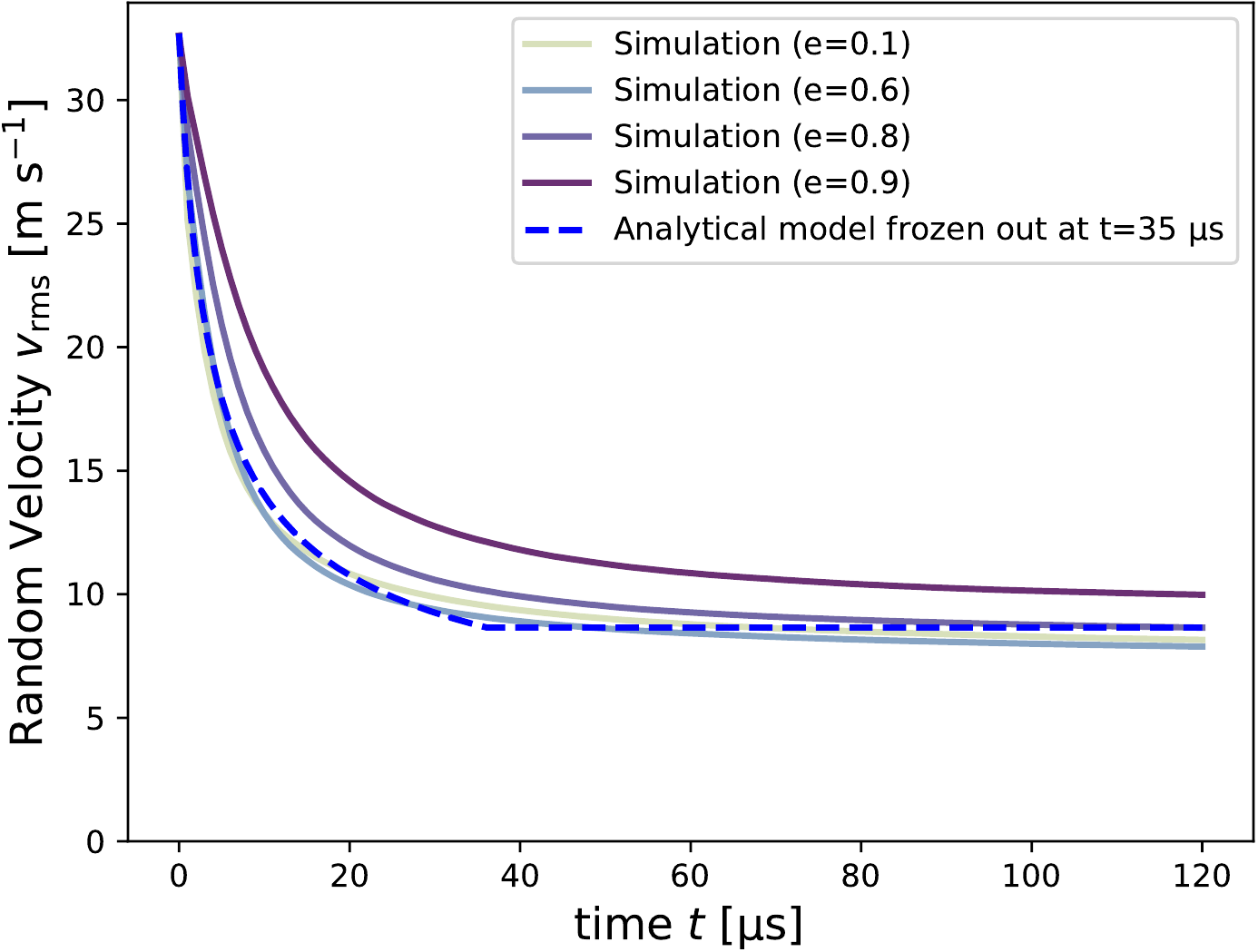}{0.45\textwidth}{}
          }
\caption{Cluster random velocity as a function of time $t$ from simulations for different values of the coefficient of restitution (solid lines), compared with the prediction from the analytic model (dashed line, same as in the left panel of Figure \ref{fig:vel}).
\label{fig:vel-restitution}}
\end{figure}
The analytic model assumes perfect coalescence of the clusters. It is natural to expect that the model would not be applicable to granular materials with a high coefficient of restitution $e$. 
To quantify the range of $e$ where the model ramains valid, 
we reran simulations presented in Section~\ref{sec:sim} but adopting different values of $e$.
Figure~\ref{fig:vel-restitution} shows the time evolution of the random velocity obtained from these simulations, compared with the prediction from the model including the cluster freezeout. 
We find that the the simulation results for $e < 0.8$ deviate from the model prediction, suggesting that the assumption of perfect coalescence is only valid for $e \la 0.8$.

The next question then is whether the values of $e$ of real granular materials falls into the range where our model remains valid.
Measurements show that $e=0.97$ for 3 mm glass beads and $e=0.87$ for 6 mm cellulose \citet{doi:10.1063/1.868282}, indicating that perfect cluster coalescence may not be a good approximation for these millimeter-sized particles. 
However, for smaller particles, cohesive forces can lead to a smaller coefficient of restitution~\citep{article}. 
There is also experimental evidence that collision velocities exceeding $10~\rm m~s^{-1}$ cause a reduction of $e$ to below 0.9 \citep{PhysRevE.56.5717}.
Therefore, we expect that the assumption of perfect cluster coalescence is valid for a high-velocity impact onto a target made of submillimeter-sized granules.

Another strong assumption in our model is that all ejecta particles are equal-sized. \citet{Kadono_2019} report that a target made of different sized granules produces ejecta curtains with filament patterns.
Extending our analytic model to polydisperse granular targets will be the subject of future work.

\subsection{Evolution of the Mesh Pattern to Crater Rays}
\label{subsec:discussion_ray}

In this section, we discuss a possible evolution path from the mesh-like pattern of an ejecta curtain to crater-rays frequently observed around flesh craters. 

Crater rays on planetary bodies have two important  features: 
(1) the rays' azimuthal pattern has a characteristic angle scale \citep{KADONO2015215}; (2) the sediments consisting of the rays are stretched in the radial direction from the host craters' center \citep[e.g.,][]{2005Icar..176..351M, KADONO2015215}. 
As already discussed by \citet{KADONO2015215} and also demonstrated in our experiment and simulations, mesh pattern formation through inelastic particle collisions followed by the pattern's geometric expansion naturally explains the first feature of the crater rays.  

\begin{figure*}
\gridline{\fig{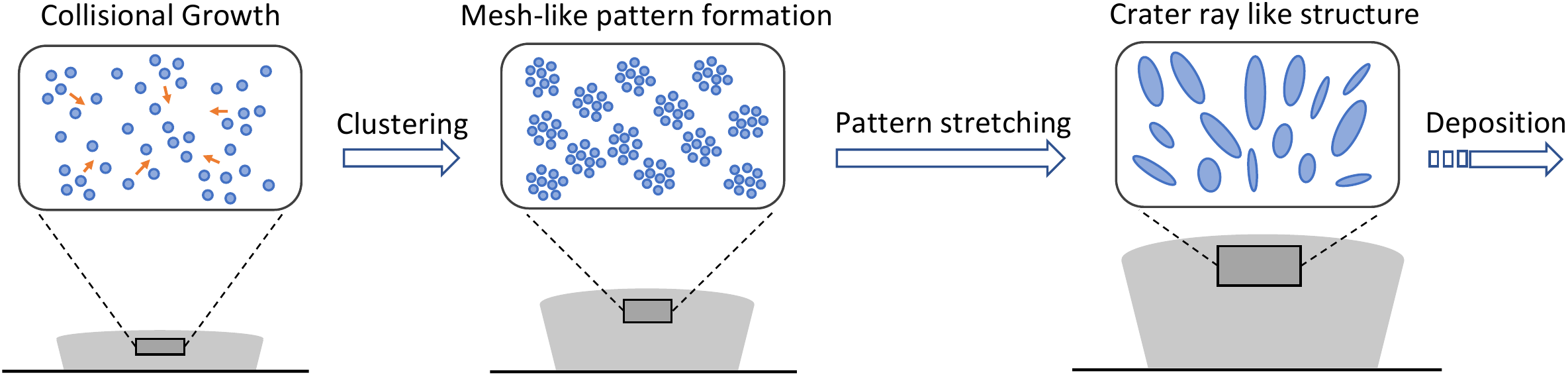}{1.0\textwidth}{}}
\caption{Illustration of a possible evolution path from the mesh pattern to the observed ray. The mesh-like pattern formed by the inelastic collision is stretched in the radial direction by the velocity contrast according to the distance from the impact point, resulting in a crater ray-like pattern. }
\label{fig:pattern_evo}
\end{figure*}
The second feature may also be explained if ejecta particles launched earlier have higher launch speeds. As we mentioned in Section~\ref{subsec:sim_result}, if the expansion associated with this velocity difference dominates over the expansion of the curtain's circumference, the particle mesh pattern should be stretched vertically. 
Indeed, it is known that the launch speed of the ejecta particles decreases with increasing distance from the impact point \citep[e.g.,][]{1983JGR....88.2485H}.
Therefore, we can envision the scenario in which the mesh pattern forming in an ejecta curtain through initial particle clustering evolves into a vertically elongated, crater ray-like pattern by the vertical expanding motion as schematically illustrated in Figure \ref{fig:pattern_evo}.
In our experiment, the pattern evolution envisioned above was not observed due to the limited time window of the high-speed imaging. On the other hand, such pattern evolution was observed in the previous impact experiments conducted by \citet{KADONO2015215}.

In the case of natural impacts occurring under low-gravity environments, such as the lunar surface, the mesh pattern is more likely to evolve into rays than in laboratory environments under the Earth's gravity due to longer flight times. 
Nevertheless, not all observed craters have rays around them. We expect that this is because of degradation.
On the Moon, the lifetime of rays is estimated to be 1 Gyr, and most of the formation ages of lunar craters are older than 1 Ga \citep{https://doi.org/10.1029/97JE00114}.
Therefore, rays are often not observed in actual craters.

\section{Summary and Conclusions}
\label{sec:conclusion}
We have conducted laboratory and numerical experiments to study pattern formation and evolution in an early phase of ejecta curtain formation.
Our key findings are summarized as follows.
\begin{enumerate}
    \item In our hyper-velocity impact experiments using a 10 mm-sized projectile, the ejecta curtain already exhibited mesh-like patterns at $10~\rm{\mu s}$ after the impact (Figure \ref{fig:experiments}). This timescale is comparable to the timescale of the ejecta's expansion, suggesting that pattern formation is completed within the expansion timescale. 
    
    \item In the numerical simulations, we employed a cluster analysis to quantify cluster formation in an expanding particle system. At early times, we find that the particle random velocity decays and the mean cluster size increases with time, confirming the prediction by \citet{KADONO2015215} that the pattern formation (particle clustering) is induced by inelastic particle collisions. At later times, the cluster pattern simply expands with time, consistent with the results of the previous laboratory experiments and ours.
    
    \item We have constructed an analytic model for pattern formation that assumes perfect coalescence of clusters at early times and collisionless geometric expansion of the clusters at later times (Section~\ref{subsec:modeling_equ}).
    The model confirms the expectation from our laboratory experiment that the initial cluster growth stage is completed on a timescale comparable to the expansion timescale of the particle system (Section~\ref{sec:freezeout}).
    Furthermore, the model reasonably reproduces the final mass of the clusters seen in our simulations as a function of initial conditions (Figure \ref{fig:vel}).
    Our model is valid as long as the coefficient of restitution of the particles is 0.8 or smaller (Figure \ref{fig:vel-restitution}).

\end{enumerate}

If the ejecta particles have a large positive vertical velocity gradient, the mesh pattern forming through inelastic particle collisions would evolve into a vertically elongated pattern. This may eventually form radially extended crater rays upon deposition (Section~\ref{subsec:discussion_ray} and  Figure~\ref{fig:pattern_evo}).
The analytic model presented in this study will allow us to better understand the pattern formation in this process.
However, our current model is limited to targets made of monodisperse particles. Extension of the model to polydisperse  granular targets should be done in future work.

\begin{acknowledgments}
We thank Takanori Iwasawa for discussions that motivated this work.
We appreciate two anonymous referees for their careful reviews that helped greatly improve the manuscript, and Dr. Edgard G. Rivera-Valentin for handling the manuscript as an editor.
This work was supported by JSPS KAKENHI Grant Numbers JP20H01948,  JP20H00182, JP19K03926, JP19K03941, JP18H05438, 17K18812 and by the Hypervelocity Impact Facility (former facility name: the Space Plasma Laboratory), ISAS, JAXA.

\software{REBOUND \citep{2012A&A...537A.128R}}
\end{acknowledgments}

\bibliographystyle{aasjournal}
\bibliography{article.bib} 

\begin{thebibliography}{}
\expandafter\ifx\csname natexlab\endcsname\relax\def\natexlab#1{#1}\fi
\providecommand{\url}[1]{\href{#1}{#1}}
\providecommand{\dodoi}[1]{doi:~\href{http://doi.org/#1}{\nolinkurl{#1}}}
\providecommand{\doeprint}[1]{\href{http://ascl.net/#1}{\nolinkurl{http://ascl.net/#1}}}
\providecommand{\doarXiv}[1]{\href{https://arxiv.org/abs/#1}{\nolinkurl{https://arxiv.org/abs/#1}}}

\bibitem[{Ester {et~al.}(1996)Ester, Kriegel, Sander, \&
  Xu}]{10.5555/3001460.3001507}
Ester, M., Kriegel, H.-P., Sander, J., \& Xu, X. 1996, in Proceedings of the
  Second International Conference on Knowledge Discovery and Data Mining,
  KDD'96 (AAAI Press), 226^^e2^^80^^93231

\bibitem[{Foerster {et~al.}(1994)Foerster, Louge, Chang, \&
  Allia}]{doi:10.1063/1.868282}
Foerster, S.~F., Louge, M.~Y., Chang, H., \& Allia, K. 1994, Physics of Fluids,
  6, 1108, \dodoi{10.1063/1.868282}

\bibitem[{Hawke {et~al.}(2004)Hawke, Blewett, Lucey, Smith, Bell, Campbell, \&
  Robinson}]{HAWKE20041}
Hawke, B., Blewett, D., Lucey, P., {et~al.} 2004, Icarus, 170, 1 ,
  \dodoi{https://doi.org/10.1016/j.icarus.2004.02.013}

\bibitem[{{Housen} {et~al.}(1983){Housen}, {Schmidt}, \&
  {Holsapple}}]{1983JGR....88.2485H}
{Housen}, K.~R., {Schmidt}, R.~M., \& {Holsapple}, K.~A. 1983, \jgr, 88, 2485,
  \dodoi{10.1029/JB088iB03p02485}

\bibitem[{{Joe} \& {Ward}(1963)}]{doi:10.1080/01621459.1963.10500845}
{Joe}, H., \& {Ward}, J. 1963, Journal of the American Statistical Association,
  58, 236, \dodoi{10.1080/01621459.1963.10500845}

\bibitem[{Kadono {et~al.}(2019)Kadono, Suetsugu, Arakawa, Kasagi, Nagayama,
  Suzuki, \& Hasegawa}]{Kadono_2019}
Kadono, T., Suetsugu, R., Arakawa, D., {et~al.} 2019, The Astrophysical
  Journal, 880, L30, \dodoi{10.3847/2041-8213/ab303f}

\bibitem[{Kadono {et~al.}(2020)Kadono, Suzuki, Matsumura, Naka, Suetsugu,
  Kurosawa, \& Hasegawa}]{KADONO2020113590}
Kadono, T., Suzuki, A.~I., Matsumura, R., {et~al.} 2020, Icarus, 339, 113590,
  \dodoi{https://doi.org/10.1016/j.icarus.2019.113590}

\bibitem[{Kadono {et~al.}(2015)Kadono, Suzuki, Wada, Mitani, Yamamoto, Arakawa,
  Sugita, Haruyama, \& Nakamura}]{KADONO2015215}
Kadono, T., Suzuki, A., Wada, K., {et~al.} 2015, Icarus, 250, 215 ,
  \dodoi{https://doi.org/10.1016/j.icarus.2014.11.030}

\bibitem[{Labous {et~al.}(1997)Labous, Rosato, \& Dave}]{PhysRevE.56.5717}
Labous, L., Rosato, A.~D., \& Dave, R.~N. 1997, Phys. Rev. E, 56, 5717,
  \dodoi{10.1103/PhysRevE.56.5717}

\bibitem[{McEwen {et~al.}(1997)McEwen, Moore, \&
  Shoemaker}]{https://doi.org/10.1029/97JE00114}
McEwen, A.~S., Moore, J.~M., \& Shoemaker, E.~M. 1997, Journal of Geophysical
  Research: Planets, 102, 9231, \dodoi{https://doi.org/10.1029/97JE00114}

\bibitem[{{McEwen} {et~al.}(2005){McEwen}, {Preblich}, {Turtle}, {Artemieva},
  {Golombek}, {Hurst}, {Kirk}, {Burr}, \& {Christensen}}]{2005Icar..176..351M}
{McEwen}, A.~S., {Preblich}, B.~S., {Turtle}, E.~P., {et~al.} 2005, \icarus,
  176, 351, \dodoi{10.1016/j.icarus.2005.02.009}

\bibitem[{{Melosh}(1989)}]{1989icgp.book.....M}
{Melosh}, H.~J. 1989, {Impact cratering : a geologic process}

\bibitem[{{Minton} {et~al.}(2019){Minton}, {Fassett}, {Hirabayashi}, {Howl}, \&
  {Richardson}}]{2019Icar..326...63M}
{Minton}, D.~A., {Fassett}, C.~I., {Hirabayashi}, M., {Howl}, B.~A., \&
  {Richardson}, J.~E. 2019, \icarus, 326, 63,
  \dodoi{10.1016/j.icarus.2019.02.021}

\bibitem[{{Oberbeck}(1971)}]{1971Moon....2..263O}
{Oberbeck}, V.~R. 1971, Moon, 2, 263, \dodoi{10.1007/BF00561880}

\bibitem[{{Rein} \& {Liu}(2012)}]{2012A&A...537A.128R}
{Rein}, H., \& {Liu}, S.~F. 2012, \aap, 537, A128,
  \dodoi{10.1051/0004-6361/201118085}

\bibitem[{{Rein} \& {Spiegel}(2015)}]{2015MNRAS.446.1424R}
{Rein}, H., \& {Spiegel}, D.~S. 2015, \mnras, 446, 1424,
  \dodoi{10.1093/mnras/stu2164}

\bibitem[{Rousseeuw(1987)}]{ROUSSEEUW198753}
Rousseeuw, P.~J. 1987, Journal of Computational and Applied Mathematics, 20, 53
  , \dodoi{https://doi.org/10.1016/0377-0427(87)90125-7}

\bibitem[{Royer {et~al.}(2009)Royer, Evans, Oyarte~G^^c3^^a1lvez, Guo, Kapit,
  M^^c3^^b6bius, Waitukaitis, \& Jaeger}]{article}
Royer, J., Evans, D., Oyarte~G^^c3^^a1lvez, L., {et~al.} 2009, Nature, 459,
  1110, \dodoi{10.1038/nature08115}

\bibitem[{{Sabuwala} {et~al.}(2018){Sabuwala}, {Butcher}, {Gioia}, \&
  {Chakraborty}}]{2018PhRvL.120z4501S}
{Sabuwala}, T., {Butcher}, C., {Gioia}, G., \& {Chakraborty}, P. 2018, \prl,
  120, 264501, \dodoi{10.1103/PhysRevLett.120.264501}

\bibitem[{{Shuvalov}(2012)}]{2012M&PS...47..262S}
{Shuvalov}, V. 2012, Meteoritics and Planetary Science, 47, 262,
  \dodoi{10.1111/j.1945-5100.2011.01324.x}

\bibitem[{{Trask} \& {Guest}(1975)}]{1975JGR....80.2461T}
{Trask}, N.~J., \& {Guest}, J.~E. 1975, \jgr, 80, 2461,
  \dodoi{10.1029/JB080i017p02461}

\end{thebibliography}

\end{document}